\documentclass[12pt]{article}
\usepackage{cite}
\newcommand{\be}{\begin{equation}}
\newcommand{\ee}{\end{equation}}
\newcommand{\bea}{\begin{eqnarray}}
\newcommand{\eea}{\end{eqnarray}}

\def\SUN{  SU($N$) }
\def\rmcp{ {\rm c.p.}}
\def\rmcyc{ {\rm cyclic~permutations}}
\def\gSYM{g_{\rm SYM}}
\def\lSYM{\lambda_{\rm SYM}}
\def\lSG{\lambda_{\rm SG}}
\def\Asy{A_{\rm SYM}}
\def\Msy{M_{\rm SYM}}
\def\Asgzero{A_{\rm SG}^{(0)}}
\def\Asgone{A_{\rm SG}^{(1)}}

\def\AsgEll{A_{\rm SG}^{(L)}}
\def\Msgone{M_{\rm SG}^{(1)}}
\def\Msgtwo{M_{\rm SG}^{(2)}}
\def\MsgEll{M_{\rm SG}^{(L)}}
\def\Tr{{\rm Tr}}
\def\e{ {\rm e} }

\newcommand{\Z}{\mathsf{Z}\kern -5pt \mathsf{Z}}
\def\half{ {1\over 2} }
\def\third{ {1\over 3} }
\def\lr{\leftrightarrow}

\def\eqn#1{eq.~(\ref{#1})} 
\def\eqns#1#2{eqs.~(\ref{#1}) and~(\ref{#2})}

\def\ket#1{|{#1}\rangle}
\def\tA{\tilde{A}}
\def\cA{  {\cal A}  }
\def\cC{  {\cal C}  }
\def\cG{  {\cal G}  }
\def\cG{  {\cal G}  }

\def\cN{  {\cal N}  }
\def\cO{  {\cal O}  }
\def\bI{\mathbf{I}}
\def\bT{\mathbf{T}}
\def\bF{\mathbf{F}}
\def\bG{\mathbf{G}}
\def\bGam{\mathbf{\Gamma}}
\def\bone{1\kern -3pt \mathrm{l}}
\def\bhH{\hat\mathbf{H}}

\def\ii#1{ {[{#1}]} }
\def\Ell{{(L)}}
\def\Ellk{{(L,k)}}
\def\Elleven{{(L,2k)}}
\def\Ellodd{{(L,2k+1)}}
\def\EllL{{(L,L)}}
\def\Ellf{{(Lf)}}
\def\EllLC{{(L,0)}}
\def\EllfLC{{(Lf,0)}}
\def\EllDT{{(L,1)}}
\def\EllSC{{(L,2)}}
\def\EllSDT{{(L,3)}}
\def\Atree{  \left( - 8 i K \over s t u \right) }
\def\Zero{{(0)}}
\def\One{{(1)}}
\def\Onef{{(1f)}}
\def\OneLC{{(1,0)}}
\def\OnefLC{{(1f,0)}}
\def\OneDT{{(1,1)}}
\def\OnefDT{{(1f,1)}}

\def\Two{{(2)}}
\def\Twok{{(2,k)}}
\def\Twof{{(2f)}}
\def\TwoLC{{(2,0)}}
\def\TwofLC{{(2f,0)}}
\def\TwoDT{{(2,1)}}
\def\TwofDT{{(2f,1)}}
\def\TwoSC{{(2,2)}}
\def\TwofSC{{(2f,2)}}
\def\TwoP{{(2)P}}
\def\TwoNP{{(2)NP}}
\def\Three{{(3)}}
\def\Threek{{(3,k)}}
\def\Threef{{(3f)}}
\def\ThreeLC{{(3,0)}}

\def\ThreeDT{{(3,1)}}

\def\ThreeSC{{(3,2)}}

\def\ThreeSDT{{(3,3)}}

\def\ep{\epsilon}
\def\cep{c(\ep)}
\def\tS{{\tt S}}
\def\tT{{\tt T}}
\def\tU{{\tt U}}
\def\ia{ \alpha_\ep}
\def\iatwo{\alpha_{2 \ep}}
\def\ib{\beta_\ep}
\def\ibtwo{\beta_{2 \ep}}
\def\ic{\gamma_\ep}
\def\ictwo{\gamma_{2\ep}}
\def\id{\delta_\ep}
\def\idtwo{\delta_{2\ep}}
\def\ha{h_\alpha}
\def\hb{h_\beta}
\def\hc{h_\gamma}
\def\hd{0}
\def\LOG{  L }

\def\ones{ \pmatrix{1\cr 1\cr 1} }

\begin{document}
\bibliographystyle{utphys}

\begin{flushright}
BRX-TH-598\\
BOW-PH-143\\
TIT/HEP-587
\end{flushright}
\vspace{25mm}

\begin{center}
{\Large\bf\sf  
Subleading-color Contributions to Gluon-gluon
Scattering in $\cN=4$ SYM Theory 
and Relations to $\cN=8$ Supergravity
} 

\vskip 5mm Stephen G. Naculich\footnote{Research supported in part by the 
NSF under grant PHY-0456944}$^{,a}$,
Horatiu Nastase,$^{b}$ 
and Howard J. Schnitzer\footnote{Research supported in part 
by the DOE under grant DE--FG02--92ER40706\\
{\tt \phantom{aaa} 
naculich@bowdoin.edu, nastase@phys.titech.ac.jp, schnitzr@brandeis.edu}
}$^{,c}$
\end{center}

\begin{center}
$^{a}${\em Department of Physics\\
Bowdoin College, Brunswick, ME 04011, USA}

\vspace{5mm}

$^{b}${\em Global Edge Institute\\
Tokyo Institute of Technology, Tokyo 152-8550, Japan} 

\vspace{5mm}

$^{c}${\em Theoretical Physics Group\\
Martin Fisher School of Physics\\
Brandeis University, Waltham, MA 02454, USA}
\end{center}
\vskip 2mm

\begin{abstract}
We study the subleading-color (nonplanar) contributions
to the four-gluon scattering amplitudes in $\cN=4$ 
supersymmetric \SUN Yang-Mills theory. 
Using the formalisms of Catani and of Sterman and Tejeda-Yeomans,
we develop explicit expressions for the infrared-divergent
contributions of all the subleading-color $L$-loop amplitudes
up to three loops,
and make some conjectures for the IR behavior for arbitrary $L$. 
We also derive several intriguing relations between
the subleading-color one- and two-loop four-gluon amplitudes 
and the four-graviton amplitudes of $\cN=8$ supergravity.
The exact one- and two-loop $\cN=8$ supergravity amplitudes 
can be expressed in terms of the 
one- and two-loop $N$-independent $\cN=4$ SYM amplitudes respectively,
but the natural generalization to higher loops fails,
despite having a simple interpretation in terms of the 't Hooft picture. 
We also find that, at least through two loops,
the subleading-color amplitudes of $\cN=4$ SYM theory
have uniform transcendentality (as do the leading-color amplitudes).
Moreover, the $\cN=4$ SYM Catani operators, which express the IR-divergent 
contributions of loop amplitudes in terms of lower-loop amplitudes,
are also shown to have uniform transcendentality,
and to be the maximum transcendentality piece of the QCD Catani operators.

\end{abstract}

\vfil\break


\section{Introduction}
\setcounter{equation}{0}
\label{secintro}

In the effort to develop new tools for the computation
of higher-loop contributions to scattering amplitudes
in gauge theories, 
$\cN=4$ SYM theory plays a special role 
because of its comparatively simple structure \cite{Anastasiou:2003kj}.
Moreover, the AdS/CFT correspondence allows some 
of the observables of this theory to be computed in the strong coupling 
limit \cite{Maldacena:1997re,Gubser:1998bc,Aharony:1999ti}.

The two-loop four-gluon scattering amplitude 
was first computed for $\cN=4$ \SUN SYM theory by 
Bern, Rosowsky, and Yan \cite{Bern:1997nh} using cutting techniques,
with the results expressed in terms of 
two-loop planar and non-planar scalar integrals 
(see also ref.~\cite{Bern:1998ug}).
Explicit expressions for these IR-divergent integrals
as Laurent expansions in $\ep$ (where $D=4-2\ep$)
were later obtained by Smirnov in the planar case \cite{Smirnov:1999gc},
and by Tausk in the non-planar case \cite{Tausk:1999vh}.
Subsequently,
Anastasiou, Bern, Dixon, and Kosower (ABDK) 
demonstrated that the two-loop amplitude  
is expressible in terms of the one-loop amplitude
in the large-$N$ (leading color) limit,
suggesting an iterative structure for the loop expansion 
of $\cN=4$ SYM amplitudes in this same limit \cite{Anastasiou:2003kj}.
Using insights from decades of study of the 
IR divergences of gauge theory amplitudes \cite{Mueller:1979ih,
Collins:1980ih, Collins:1989bt, Sen:1981sd, Magnea:1990zb, Giele:1991vf,
Kunszt:1994mc, Catani:1996vz, Dixon:2008gr, Catani:1998bh, Sterman:2002qn},
Bern, Dixon, and Smirnov 
conjectured a complete nonperturbative exponential ansatz 
for MHV $n$-gluon scattering amplitudes \cite{Bern:2005iz},
again in the leading-color (large-$N$) limit. 
In this limit, only planar diagrams 
(in the topological expansion of 't Hooft) contribute \cite{'tHooft:1973jz}.

In this paper, we explore the structure of the subleading-color
(nonplanar) contributions to the four-gluon amplitude
in $\cN=4$  \SUN     SYM theory through three loops,
focusing particularly on the IR-divergent terms.
We also demonstrate some intriguing connections between these
subleading-color amplitudes and
four-graviton amplitudes in $\cN=8$ supergravity.

After reviewing the known exact one- and two-loop results for the
full four-gluon amplitude in $\cN=4$ SYM theory,
we use Catani's formalism \cite{Catani:1998bh} 
to develop explicit expressions for the IR-divergent terms 
of the one- and two-loop subleading-color amplitudes,
and a combination of his approach and that of 
Sterman and Tejeda-Yeomans \cite{Sterman:2002qn}
for three loops. 
We denote the leading-color $L$-loop amplitude by $A^\EllLC$,
and the subleading-color amplitudes by $A^\Ellk$,
($k=1, \cdots, L$)
where $A^\Ellk$ is the component of the amplitude 
proportional to $N^{L-k}$.
We show (for $L\le 3$)
that the leading IR divergence of the amplitude 
$A^\Ellk$ is $\cO(1/\ep^{2L-k})$, 
and explicitly determine its coefficient.
We find that the first two terms of the Laurent expansion of 
the $N$-independent amplitude $A^\EllL$ 
(which starts at $\cO(1/\ep^L)$) 
obey the relationship
\be
A^\EllL (\ep) \quad \sim \quad
\frac{ P_{L-1} (X,Y,Z) }{\ep^{L-1}} A^\OneDT (L \ep)
\, + \, \cO \left(1 \over \ep^{L-2} \right)
\label{introLL}
\ee
where $P_n(X,Y,Z)$ is an $n$th order polynomial,
explicitly specified in \eqns{Lodd}{Leven},
and 
$X = \log \left(t /u\right)$,
$Y = \log \left(u / s\right)$,
and $Z = \log \left(s / t\right)$,
with $s$, $t$, and $u$ being Mandelstam parameters.
We prove \eqn{introLL} for two and three loops
(cf. \eqns{twoloopmat}{threeeps}),
and conjecture the result to be valid generally. 

Next,
observing that the $N$-independent SYM amplitude $A^\EllL$
has a leading divergence of the same order as the 
$L$-loop four-graviton 
$\cN=8$ supergravity amplitude \cite{Naculich:2008ew,Brandhuber:2008tf},
we show in sec.~\ref{secSG} that the full amplitudes
are related.
In particular, we demonstrate the exact relationships
for one- and two-loop amplitudes 
\bea
\frac{1}{3}\left[(\lSG u)\Msy^\OneDT(s,t) + \rmcyc \right]
&=&\sqrt2\lSYM\Msgone
\label{symsgv}
\\
\frac{1}{3}
\left[(\lSG u)^2\Msy^\TwoSC (s,t) + \rmcyc \right]
&=&
(\sqrt2\lSYM)^2 \Msgtwo
\label{symsgvtwo}
\eea
where $\Msy^\EllL$ and $\MsgEll$  are ratios of 
$L$-loop $N$-independent SYM amplitudes $\Asy^\EllL$ and
$L$-loop supergravity amplitudes  $\AsgEll$
to tree-level amplitudes,
and $\lSYM= g^2 N$ and $\lSG = (\kappa/2)^2 $ 
are SYM and supergravity coupling constants.
The natural generalization of \eqns{symsgv}{symsgvtwo}
to $L$ loops is not satisfied,
at least in its simplest form.
The one- and two-loop relations have a simple (albeit nonintuitive)
interpretation in terms of the 't Hooft picture, 
once we factor in an unusual identification, in that 
the topological expansion of the SYM Feynman diagrams is 
related to the loop expansion of supergravity. 

In sec.~\ref{sectrans},
we discuss the transcendentality of the $\cN=4$ SYM amplitudes
as well as that of the Catani {\it operators}, which
has not previously appeared in the literature.
We observe that the subleading-color $\cN=4$ SYM amplitudes 
through two loops have uniform transcendentality,
as is already known for the leading-color amplitudes.
Moreover, the $\cN=4$ SYM Catani operators (at least through three loops)
also have uniform transcendentality,
which implies the same for the divergent contributions
of $n$-gluon scattering amplitudes through that loop order. 
Finally, the $\cN=4$ SYM Catani operators constitute
the maximum transcendentality piece of the corresponding QCD 
Catani operators.

\section{One- and two-loop four-gluon amplitudes}  
\setcounter{equation}{0}
\label{secfourgluon} 

In this section, we review known exact results for one- and two-loop
four-gluon amplitudes in $\cN=4$ \SUN SYM theory, 
and relations among them.

Gluon $n$-point amplitudes may be written in a color-decomposed form 
as a sum over single and multiple traces of color generators \cite{Bern:1990ux}.
The four-gluon amplitude contains only single and double traces, 
and can be written as 
\bea
{\cal A}_{4-{\rm gluon}}(1,2,3,4)
&=&
\sum_{\sigma\in S_4/\Z_4}
\Tr(T^{a_{\sigma(1)}}T^{a_{\sigma(2)}}T^{a_{\sigma(3)}} T^{a_{\sigma(4)}})
A_{4;1}(\sigma(1),\sigma(2),\sigma(3),\sigma(4))
\nonumber\\
&+&\sum_{\sigma\in S_4/\Z_2^2}
\Tr(T^{a_{\sigma(1)}}T^{a_{\sigma(2)}})
\Tr(T^{a_{\sigma(3)}}T^{a_{\sigma(4)}})
A_{4;3}(\sigma(1),\sigma(2), \sigma(3),\sigma(4))
\nonumber\\
\label{colordecomp}
\eea
where the color-stripped amplitudes $A_{4;1}$  and $A_{4;3}$
implicitly depend on the momenta and polarizations of the external particles,
and where $T^a$ are SU$(N)$ generators in the fundamental representation,
normalized according to $\Tr (T^a T^b) = \delta^{ab}$.
The $L$-loop diagrams contributing to $A_{4;1}$ start at order $N^L$, 
those contributing to $A_{4;3}$ start at order $N^{L-1}$, 
and corrections to each term come with factors of $1/N^2$ 
(from loop index traces), 
so that the $L$-loop amplitude has the form 
\bea
A_{4;1} &=&
g^2 a^L \left[A_{4;1}^{\EllLC} +\frac{1}{N^2}A_{4;1}^{\EllSC}+ \cdots\right]
\nonumber\\
A_{4;3} &=& g^2 a^L \left[
 \frac{1}{N}A_{4;3}^{\EllDT}+\frac{1}{N^3}A_{4;3}^{\EllSDT}+ \cdots \right]
\eea
with the series ending at the $N$-independent amplitude $A^\EllL$, 
and where the natural 't Hooft loop expansion parameter is \cite{Bern:2005iz}
\be
a \equiv {g^2 N \over 8 \pi^2} \left( 4 \pi \e^{-\gamma} \right)^\ep \,.
\ee
Here $\gamma$ is Euler's constant,
and the loop amplitudes, being IR-divergent, are 
evaluated using dimensional regularization in 
$D= 4 - 2 \ep$ dimensions.
The leading-color term $A^\EllLC$ comes from planar diagrams, 
whereas the subleading-color terms $A^\EllDT$ through $A^\EllL$ 
include contributions from nonplanar diagrams.

We will also find it convenient to write the 
four-gluon amplitude as
\bea
\cA_{4-{\rm gluon}} (1,2,3,4)
&=&  g^2 \sum_{L=0}^\infty a^L  \sum_{i=1}^{9} A^\Ell_\ii{i}  \,\, \cC_\ii{i}
\nonumber\\
&=& 
 g^2 \sum_{L=0}^\infty a^L  \sum_{k=0}^L \frac{1}{N^k} \sum_{i=1}^{9}
A^\Ellk_\ii{i}  \,\, \cC_\ii{i}
\label{ampEll}
\eea
in terms of an explicit basis of traces  \cite{Glover:2001af} 
\bea
\label{basis}
&& \hspace{-5mm}
   \cC_\ii{1} = \Tr(T^{a_1} T^{a_2} T^{a_3} T^{a_4})\,, \qquad
   \cC_\ii{4} = \Tr(T^{a_1} T^{a_3} T^{a_2} T^{a_4})\,, \qquad
   \cC_\ii{7} = \Tr(T^{a_1} T^{a_2}) \Tr(T^{a_3} T^{a_4}) \nonumber\\
&& \hspace{-5mm}
   \cC_\ii{2} = \Tr(T^{a_1} T^{a_2} T^{a_4} T^{a_3})\,, \qquad
   \cC_\ii{5} = \Tr(T^{a_1} T^{a_3} T^{a_4} T^{a_2})\,, \qquad
   \cC_\ii{8} = \Tr(T^{a_1} T^{a_3}) \Tr(T^{a_2} T^{a_4}) \nonumber\\
&& \hspace{-5mm}
   \cC_\ii{3} = \Tr(T^{a_1} T^{a_4} T^{a_2} T^{a_3})\,, \qquad
   \cC_\ii{6} = \Tr(T^{a_1} T^{a_4} T^{a_3} T^{a_2})\,, \qquad
   \cC_\ii{9} = \Tr(T^{a_1} T^{a_4}) \Tr(T^{a_2} T^{a_3})  \nonumber\\
\eea
so that $A^\Elleven_\ii{1}$ through $A^\Elleven_\ii{6}$ 
correspond to the single-trace amplitudes $A^\Elleven_{4;1}$, 
and $A^\Ellodd_\ii{7}$ through $A^\Ellodd_\ii{9}$ 
correspond to the double-trace amplitudes $A^\Ellodd_{4;3}$.

The tree-level amplitudes are 
\be
\left( 
A^\Zero_\ii{1},  \,
A^\Zero_\ii{2}, \,
A^\Zero_\ii{3}, \,
A^\Zero_\ii{4},  \,
A^\Zero_\ii{5},  \,
A^\Zero_\ii{6} \right)
= 
- \frac{ 4 i K }{stu}
\left(u, t, s, s, t, u\right)
\label{tree}
\ee 
where $s=(k_1+k_2)^2$, $t=(k_1+k_4)^2$,  and $u=(k_1+k_3)^2$
are the usual Mandelstam variables,
obeying $s+t+u=0$ for massless external gluons.
The factor $K$, defined in eq.~(7.4.42) of ref.~\cite{Green:1987sp},
depends on the momenta and helicity of the external gluons, 
and is totally symmetric under permutations of the external legs.
The identities 
$A^\Ell_\ii{1} = A^\Ell_\ii{6}$,
$A^\Ell_\ii{2} = A^\Ell_\ii{5}$,
and 
$A^\Ell_\ii{3} = A^\Ell_\ii{4}$
are satisfied at all loop orders. 

At one loop, the single-trace amplitudes are given by \cite{Green:1982sw}
\be
\label{oneloopST}
A^\OneLC_\ii{1} 
=  M^\One (s,t) \, A^\Zero_\ii{1}
=  2 i K \, I_4^\One(s,t)
\ee
with the other 
single-trace amplitudes $A^\OneLC_\ii{2}$ and $A^\OneLC_\ii{3}$
obtained by letting $t \lr u$ and $s \lr u$ respectively. 
In eq.~(\ref{oneloopST}),
$I_4^\One (s,t)$ denotes the scalar box integral
\bea
M^\One (s,t) &=& - \half st\, I_4^\One (s,t)
\label{Monedef}
\\
I_4^\One (s,t)
= I_4^\One (t,s)
&=& -i \mu^{2\ep} \e^{\ep \gamma} \pi^{-D/2}  \int 
{d^D p \over p^2 (p-k_1)^2 (p-k_1-k_2)^2 (p+k_4)^2 } 
\nonumber
\eea
an explicit expression for which is given, e.g., in ref.~\cite{Bern:2005iz}.
The one-loop double-trace amplitudes are given by \cite{Green:1982sw}
\bea
A^\OneDT_\ii{7} =
A^\OneDT_\ii{8} =
A^\OneDT_\ii{9} 
&=&
2 \left( A^\OneLC_\ii{1} + A^\OneLC_\ii{2} + A^\OneLC_\ii{3}  \right)
\label{oneloopdecouple}
\\
&=&
4 i K \left[ I_4^\One (s,t)+ I_4^\One (t,u)+ I_4^\One (u,s) \right] \,.
\label{oneloopnp}
\eea
The relation (\ref{oneloopdecouple}) follows from the 
one-loop U(1) decoupling identity \cite{Bern:1990ux}.

At two loops, the leading-color single-trace amplitude 
is given by  \cite{Bern:1997nh}
\be
\label{twoloopLC}
A_\ii{1}^\TwoLC 
= M^\Two (s,t) \, A^\Zero_\ii{1}
= - i K \left[ s I_4^\TwoP (s,t) +t I_4^\TwoP (t,s) \right]
\ee
where $I_4^\TwoP(s,t)$ denotes the scalar double-box (planar) integral 
\bea
M^\Two(s,t) &=&
  \frac{1}{4} st\, \left[ s I_4^\TwoP (s,t) +t I_4^\TwoP (t,s) \right]
\label{Mtwodef}
\\
 I_4^\TwoP(s,t) &=&
\left( -i \mu^{2\ep} \e^{\ep \gamma} \pi^{-D/2} \right)^2
\int 
 {d^D p \, d^D q
\over p^2 \, (p + q)^2 q^2 \, (p - k_1)^2 \,(p - k_1 - k_2)^2 \,
        (q-k_4)^2 \, (q - k_3 - k_4)^2 } 
\nonumber
\eea
an explicit expression for which is given, e.g., in ref.~\cite{Bern:2005iz}.
The double-trace amplitude is \cite{Bern:1997nh}
\bea
\hspace{-5mm}
A_\ii{7}^\TwoDT &=& 
-2 i K
\Bigl[
s \left( 3 I_4^\TwoP(s, t) + 2 I_4^\TwoNP(s, t)
+ 3 I_4^\TwoP(s, u)  + 2 I_4^\TwoNP(s, u)\right)
\label{twoloopDT}
\\ && \hspace{11mm}
-t \left(I_4^\TwoNP(t, s) + I_4^\TwoNP(t, u)\right)
-u \left(I_4^\TwoNP(u, s) + I_4^\TwoNP(u,t)\right)
\Bigr]  
\nonumber
\eea
and the subleading-color single-trace amplitude is \cite{Bern:1997nh}
\bea
\hspace{-9mm}
A_\ii{1}^\TwoSC &=& 
-2 i K \,
\Bigl[
s \left(I_4^\TwoP(s, t) + I_4^\TwoNP(s, t)
+ I_4^\TwoP(s, u)  + I_4^\TwoNP(s, u)\right)
\label{twoloopSC}
\\ && \hspace{11mm}
+t \left(I_4^\TwoP(t, s) + I_4^\TwoNP(t, s)
+ I_4^\TwoP(t, u)  + I_4^\TwoNP(t, u)\right)
\nonumber\\ && \hspace{10mm}
-2u \left(I_4^\TwoP(u, s) + I_4^\TwoNP(u, s)
+ I_4^\TwoP(u,t)  + I_4^\TwoNP(u,t)\right)
\Bigr]  
\nonumber
\eea
where $I_4^\TwoNP(s,t)$ denotes the two-loop non-planar integral
\begin{equation} 
I_4^\TwoNP(s,t)  = 
\left( -i \mu^{2\ep} \e^{\ep \gamma} \pi^{-D/2} \right)^2 \int 
{d^D p \, d^D q\over p^2\,(p+q)^2\, q^2 \, (p-k_2)^2  \,(p+q+k_1)^2\,
  (q-k_3)^2 \, (q-k_3-k_4)^2}  
\label{twoloopNP}
\ee
an explicit expression for which is given in ref.~\cite{Tausk:1999vh}.
All the other single- and double-trace amplitudes $A^\Twok_\ii{i}$ 
are obtained by making the appropriate permutations of $s$, $t$, and $u$ 
in these expressions.

The two-loop amplitudes obey the following group theory 
relations \cite{Bern:2002tk}
\bea
A_\ii{7}^\TwoDT &=& 
2 \left( A^\TwoLC_\ii{1} + A^\TwoLC_\ii{2} + A^\TwoLC_\ii{3}  \right)
- A_\ii{3}^\TwoSC \nonumber\\
A_\ii{8}^\TwoDT &=& 
2 \left( A^\TwoLC_\ii{1} + A^\TwoLC_\ii{2} + A^\TwoLC_\ii{3}  \right)
- A_\ii{1}^\TwoSC \nonumber\\
A_\ii{9}^\TwoDT &=& 
2 \left( A^\TwoLC_\ii{1} + A^\TwoLC_\ii{2} + A^\TwoLC_\ii{3}  \right)
- A_\ii{2}^\TwoSC
\label{STDT}
\eea
and may be easily verified using 
eqs.~(\ref{twoloopLC}), 
(\ref{twoloopDT}), 
and (\ref{twoloopSC}).
In addition, we have
\be
\label{sumSCvanishes}
A^\TwoSC_\ii{1} + A^\TwoSC_\ii{2} + A^\TwoSC_\ii{3} =0
\ee
also easily verified using \eqn{twoloopSC}.
Together these equations imply
\be
6 \, \sum_{i=1}^3   A_\ii{i}^\TwoLC -
\sum_{i=7}^9   A_\ii{i}^\TwoDT =0
\label{twoloopdecouple}
\ee
which is the two-loop generalization of the U(1) decoupling 
relation (\ref{oneloopdecouple}).
Both \eqns{sumSCvanishes}{twoloopdecouple} are encapsulated in the
equation
\be
6\, \sum_{i=1}^3   A_\ii{i}^\Ell -
N \,\sum_{i=7}^9   A_\ii{i}^\Ell =0, \qquad L \le 2
\ee
which is valid through two loops.

\section{IR divergences of $\cN=4$ SYM amplitudes}
\setcounter{equation}{0}
\label{secIR}

When we dimensionally regularize a theory in $D=4 - 2 \ep$ dimensions,
both UV and IR divergences appear as poles in $\ep$.
In a UV finite theory, such as $\cN=4$ SYM, the
poles in $\ep$ are solely due to IR divergences.
In gluon-gluon scattering in ${\cN}=4$ SYM, 
IR divergences arise both from soft gluons 
and from collinear gluons (which can exchange a virtual gluon
with soft transverse momentum), 
each of which gives rise to an $\cO(1/\ep)$ pole at 1-loop, 
leading to an $\cO(1/\ep^2)$ pole at that order. 
At $L$ loops, the leading IR divergence of $A^\EllLC$ 
is therefore $\cO(1/\ep^{2L})$, 
arising from multiple soft gluon exchanges. 
The IR divergences of subleading-color amplitudes $A^\Ellk$, however,
are not so severe, being only $\cO(1/\ep^{2L-k})$.
In particular, 
the $N$-independent amplitude $A^\EllL$
has a leading $\cO(1/\ep^L)$ divergence,
the same degree of IR divergence as an $L$-loop
$\cN=8$ supergravity amplitude.
As we will see in sec.~\ref{secSG},
there are some intriguing connections between 
the $N$-independent $\cN=4$ SYM amplitude $A^\EllL$ 
and the $L$-loop $\cN=8$ supergravity amplitude.

In this section and the next, we will analyze the 
IR-divergent contributions of leading- and subleading-color 
$\cN=4$ SYM amplitudes up to three loops
using the general analysis of 
refs.~\cite{Catani:1998bh,Sterman:2002qn}.
This will allow us to illustrate the IR behavior described above
as well as 
to derive some relations for subleading-color amplitudes.
In this section, 
it will be useful to organize the color-stripped amplitudes
$A_\ii{i}^\Ell$ defined in \eqn{ampEll}
into a vector
\be
\ket{A^\Ell}=\left(
A^\Ell_\ii{1},  \,
A^\Ell_\ii{2}, \,
A^\Ell_\ii{3}, \,
A^\Ell_\ii{4},  \,
A^\Ell_\ii{5},  \,
A^\Ell_\ii{6},  \,
A^\Ell_\ii{7}, \, 
A^\Ell_\ii{8}, \, 
A^\Ell_\ii{9}  \right)^T
\label{ketEll}
\ee
where $( \cdots )^T$ denotes the transposed vector,
so that the loop expansion of the four-gluon amplitude 
(\ref{ampEll}) may be expressed as 
\be
\cA_{4-{\rm gluon}} 
= g^2 \Biggl [  \ket{A^\Zero}
+ a \ket{A^\One} + a^2 \ket{A^\Two} + a^3 \ket{A^\Three} + \cdots
 \Biggr ]
,\qquad
a \equiv {g^2 N \over 8 \pi^2} \left( 4 \pi \e^{-\gamma} \right)^\ep  \,.
\label{expansion}
\ee

\subsection{Catani's Formalism}

In ref.~\cite{Catani:1998bh}, 
Catani showed that the IR divergences of the $L$-loop amplitude
$\ket{A^\Ell (\ep)}$
can be characterized in terms of operators $\bI^\Ell$
acting on lower-order terms in the loop expansion
\bea
\ket{A^\One (\ep)} 
&=& \frac{1}{N} \bI^\One(\ep) \ket{A^\Zero} + \ket{A^\Onef (\ep)} 
\label{oneloopCatani}
\\
\ket{A^\Two (\ep)} 
&=& \frac{1}{N^2} \bI^\Two(\ep) \ket{A^\Zero}  
+ \frac{1}{N} \bI^\One(\ep)\ket{A^\One (\ep)} 
+ \ket{A^\Twof (\ep)}  
\label{twoloopCatani}
\\
\ket{A^\Three (\ep)} 
&=& \frac{1}{N^3} \bI^\Three(\ep) \ket{A^\Zero}  
 +  \frac{1}{N^2} \bI^\Two(\ep) \ket{A^\One}  
+ \frac{1}{N} \bI^\One(\ep)\ket{A^\Two (\ep)} 
+ \ket{A^\Threef (\ep)}  
\label{threeloopCatani}
\eea
where $\bI^\Ell (\ep)$ contains the terms that diverge as $\ep \to 0$,
and $\ket{A^\Ellf (\ep)}$ is finite as $\ep\to0$
(but is not the entire finite piece of $\ket{A^\Ell(\ep)}$, 
since $\bI^\Ell(\ep)$ contains finite terms as well).

If we specialize to four-gluon scattering in $\cN=4$ SYM theory
(for which the $\beta$-function vanishes),
the one-loop Catani operator $\bI^\One(\ep)$ takes the form\footnote{We 
adopt the normalization convention of ref.~\cite{Bern:2005iz}, 
omitting the prefactor $\e^{\ep \gamma}/\Gamma(1-\ep)$  that appears
in refs.~\cite{Catani:1998bh,Glover:2001af,Bern:2002tk,Bern:2004cz}.
This only affects the form of the finite contribution $\ket{A^\Ellf (\ep)}$.}
\be 
\bI^\One(\ep) = {1\over2\ep^2} 
 \sum_{i=1}^4 \sum_{j\neq i}^4 \bT_i \cdot \bT_j
 \left( { \mu^2 \over -s_{ij} } \right)^{\ep} 
\label{Ionedef}
\ee 
where $\bT_i \cdot \bT_j = T_i^a T_j^a$ 
and $T_i^a$ are the SU$(N)$ generators in 
the adjoint representation.

The two-loop Catani operator $\bI^\Two(\ep)$ may be
written in the case of $\cN=4$ SYM as 
\cite{Bern:2002tk,Bern:2004cz}
\be
\bI^\Two(\ep) 
=
- {1\over2} \left[ \bI^\One(\ep) \right]^2
- N \zeta_2 
\, \cep \, 
\, \bI^\One(2\ep) 
+ 
\frac{\cep}{4 \ep} 
  \biggl[ 
  - \frac{N \zeta_3}{2} \sum_{i=1}^4 \sum_{j\neq i}^4 \, \bT_i \cdot \bT_j
    \Bigl( { \mu^2 \over -s_{ij} } \Bigr)^{2\ep}
       + \bhH^\Two \biggr] 
\label{Itwodef}
\ee
where
\bea
\cep \equiv \e^{-\ep\gamma} \Gamma(1-\ep)
&=&
\frac{\pi \ep}{\sin(\pi\ep)} \exp 
\left[\sum_{k=2}^\infty (-1)^{k+1} \frac{\zeta_k \ep^k}{k} \right]
= 1+ \frac{\pi^2}{12}\,\ep^2   + \cO(\ep^3)
\label{cep} \\
\bhH^\Two 
&=& 
- 4 \LOG  \left[ \bT_1 \cdot \bT_2, \bT_2 \cdot \bT_3 \right]
\label{Hhatdef} \\ [2mm] 
\LOG 
&=& 
\log \left(s\over t\right) 
\log \left(t\over u\right) \log \left(u\over s\right)\,.
\label{triplelog}
\eea
We may use \eqn{Ionedef} to rewrite \eqn{Itwodef} as
\be
\bI^\Two(\ep) 
= - {1\over2} \left[ \bI^\One(\ep) \right]^2
- N (\zeta_2 + \ep \zeta_3)  
\, \cep \, 
\bI^\One(2\ep) 
+ 
\frac{\cep}{4 \ep} 
\bhH^\Two  \,.
\label{Itwomatrix}
\ee 
Using \eqn{Itwomatrix}, 
\eqn{twoloopCatani} may be rewritten in the form
\bea
\ket{A^\Two (\ep)} 
&=& 
  \frac{1}{2 N} \bI^\One(\ep)\ket{A^\One (\ep)} 
- \frac{1}{N} (\zeta_2 + \ep \zeta_3)  
\, \cep \, \bI^\One(2\ep) \ket{A^\Zero}  
\nonumber\\
&&+ \frac{1}{4 N^2} 
\frac{\cep}{\ep} 
\bhH^\Two \ket{A^\Zero}  
+ \frac{1}{2 N} \bI^\One(\ep)\ket{A^\Onef(\ep)} 
+ \ket{A^\Twof (\ep)}  
\label{twoloopCataniagain}
\eea
which will be useful in sec.~\ref{secexpansion}.

To determine the form of the three-loop Catani operator $\bI^\Three$, we 
turn now to the slightly different IR analysis of
Sterman and Tejeda-Yeomans \cite{Sterman:2002qn}.

\subsection{Formalism of Sterman and Tejeda-Yeomans}

In ref.~\cite{Sterman:2002qn}, 
Sterman and Tejeda-Yeomans
characterized the IR divergences of the $L$-loop amplitude
$\ket{A^\Ell (\ep)}$ as 
\bea
\ket{A^\One (\ep)} 
&=& \frac{1}{N} \bF^\One(\ep) \ket{A^\Zero} + \ket{\tA^\Onef (\ep)}
\label{oneloopSTY}
\\
\ket{A^\Two (\ep)} 
&=& \frac{1}{N^2} \bF^\Two(\ep) \ket{A^\Zero}  
+ \frac{1}{N} \bF^\One(\ep)\ket{A^\One (\ep)} 
+ \ket{\tA^\Twof (\ep)}  
\label{twoloopSTY}
\\
\ket{A^\Three (\ep)} 
&=& \frac{1}{N^3} \bF^\Three(\ep) \ket{A^\Zero}  
 +  \frac{1}{N^2} \bF^\Two(\ep) \ket{A^\One}  
+ \frac{1}{N} \bF^\One(\ep)\ket{A^\Two (\ep)} 
+ \ket{\tA^\Threef (\ep)}  
\label{threeloopSTY}
\eea
where we have rescaled\footnote{Note also that the expansion in
ref.~\cite{Sterman:2002qn} is in powers of $\alpha/\pi$ whereas
in eq.~(\ref{expansion}) the expansion is in powers of $a$.  
The only effect of this difference on the equations, 
however, is to change the numerical values of the constants 
$\gamma^\Ell$ and $\cG_0^\Ell$ in eq.~(\ref{Gdef}).}
the operators $\bF^\Ell$ of ref.~\cite{Sterman:2002qn}
by a factor of $N^L$.
The operators $\bF^\Ell (\ep)$ differ from the Catani operators 
$\bI^\Ell (\ep)$
introduced in the previous subsection
in that they contain only the divergent terms of the expansion in 
$\ep$ whereas the expansion of $\bI^\Ell (\ep)$ also contains 
non-negative powers of $\ep$.
For this reason, the finite pieces $\ket{\tA^\Ellf (\ep)}$
of eqs.~(\ref{oneloopSTY})--(\ref{threeloopSTY}) 
differ from the $\ket{A^\Ellf (\ep)}$
of eqs.~(\ref{oneloopCatani})--(\ref{threeloopCatani}), 
which is why we have distinguished them with a tilde.

Specializing to the case of $gg \to gg$ in $\cN=4$ SYM theory, 
we may rewrite the expressions for  $\bF^\Ell (\ep)$
given in ref.~\cite{Sterman:2002qn} as
\bea
\bF^{\One}(\ep)
&=&  \bG^\One (\ep)
\label{Fonedef}\\
\bF^{\Two}(\ep)
&=&
- \half \left[ \bF^\One (\ep) \right]^2 + \bG^\Two (2 \ep)
\label{Ftwodef}\\
\bF^{\Three}(\ep)
&=&
- \third \left[ \bF^\One (\ep) \right]^3 
- \third \bF^\One (\ep) \bF^\Two (\ep)
- \frac{2}{3} \bF^\Two (\ep) \bF^\One (\ep) 
+ \bG^\Three (3 \ep)
\label{Fthreedef}
\eea
with
\be
\bG^\Ell (\ep) =
\frac{N^L}{2}
\left[-\left(
\frac{\gamma^\Ell}{\ep^2}
+\frac{2\cG_0^\Ell} {\ep}  \right)\bone 
+\frac{1}{\ep}\bGam^\Ell\right]
\label{Gdef}
\ee
where $\bGam^\Ell$ are nontrivial anomalous dimension matrices and 
$\gamma^\Ell$ and $\cG_0^\Ell$ are the coefficients of 
the soft (or Wilson line cusp) and collinear anomalous dimensions 
of the gluon respectively, 
which are just numbers (proportional to the identity matrix)
\bea
\gamma (a)
&=& 
\sum_{L=1}^\infty  \gamma^\Ell a^L    = 
4 a - 4 \zeta_2 a^2 + 22 \zeta_4 a^3 + \cdots
\nonumber\\
\cG_0(a)
&=& 
\sum_{L=1}^\infty  
\cG_0^\Ell a^L  =
    -  \zeta_3 a^2 + (4 \zeta_5 + \frac{10}{3} \zeta_2 \zeta_3 )  a^3 + \cdots
\eea
These values were calculated in ref.~\cite{Bern:2005iz} 
in the planar (leading $N$) limit, 
but they remain valid for arbitrary $N$ 
because contributions subleading-in-$1/N$ 
are never proportional to the identity, 
and for $\bG^\Ell (\ep)$ 
contribute only to the anomalous dimension matrices $\bGam^\Ell$.
Moreover, $\gamma(a)$ and $\cG_0(a)$ 
are the leading transcendentality part of the corresponding 
QCD anomalous dimensions, as one can easily check 
using the formulas in ref.~\cite{Sterman:2002qn}. 

\subsection{Comparison of IR formalisms}

We now show that the operators defined by Catani and by
Sterman/Tejeda-Yeomans
are related by the following equations
\bea
\bF^\One (\ep)
&=&
\bI^\One (\ep) - \Delta^\One(\ep)
\label{FIone}\\
\bF^\Two (\ep)
&=&
\bI^\Two (\ep) + \Delta^\One(\ep) \, \bI^\One (\ep) 
- \Delta^\Two(\ep)
\label{FItwo}\\
\bF^\Three (\ep)
&=&
\bI^\Three(\ep) + \Delta^\One(\ep) \, \bI^\Two (\ep) 
+ \Delta^\Two(\ep) \, \bI^\One (\ep) 
- \Delta^\Three(\ep)
\label{FIthree}
\eea
and 
\bea
\ket{\tA^\Onef (\ep)}  
&=&
\ket{A^\Onef (\ep)}  
+ \frac{1}{N} \Delta^\One(\ep) \ket{A^\Zero}
\nonumber
\\
\ket{\tA^\Twof (\ep)}  
&=&
\ket{A^\Twof (\ep)}  
+  \frac{1}{N}\Delta^\One(\ep) \ket{A^\Onef (\ep)}  
+ \frac{1}{N^2} \Delta^\Two(\ep) \ket{A^\Zero}
\\
\ket{\tA^\Threef (\ep)}  
&=&
\ket{A^\Threef (\ep)}  
+  \frac{1}{N} \Delta^\One(\ep) \ket{A^\Twof (\ep)}  
+ \frac{1}{N^2} \Delta^\Two(\ep) \ket{A^\Onef (\ep)}  
+ \frac{1}{N^3}  \Delta^\Three(\ep) \ket{A^\Zero}
\nonumber
\eea
where the $\Delta^\Ell(\ep)$ are finite as $\ep \to 0$.

By comparing \eqn{Ionedef} with \eqn{Fonedef}, one  may ascertain
that the $\cO(1/\ep^2)$ terms on both sides of eq.~(\ref{FIone})  
match because 
$\sum_{i=1}^4 \sum_{j\neq i}^4 \bT_i \cdot \bT_j = -4 N\bone$.
The $\cO(1/\ep)$ terms match provided $\bGam^\One$ is given by 
\be
\bGam^\One = \frac{1}{N}  \sum_{i=1}^4 \sum_{j\neq i}^4 \bT_i \cdot \bT_j
 \log \left( {\mu^2 \over -s_{ij}  } \right) \,.
\ee 
The remaining (finite) part of $\bI^\One(\ep)$  defines $\Delta^\One(\ep)$,
the first term of which is
\be
\Delta^\One(\ep) = \frac{1}{4} \sum_{i=1}^4 \sum_{j\neq i}^4 \bT_i \cdot \bT_j
 \log^2  \left( {\mu^2 \over -s_{ij}  } \right) + \cO(\ep)\,.
\ee

Next by using \eqn{FItwo}, we see that
\eqns{Itwomatrix}{Ftwodef} are compatible provided that\footnote{
Further calculation shows \cite{MertAybat:2006wq, MertAybat:2006mz}
that $\bGam^\Two = - \zeta_2 \bGam^\One$,
since the last two terms of \eqn{identif} exactly cancel, 
as can be seen using eqns.~(D5)-(D8) of ref.~\cite{MertAybat:2006mz}.
We thank Lance Dixon for pointing this out.}
\be
\bGam^\Two = - \zeta_2 \bGam^\One + \frac{1}{N^2} \bhH^\Two 
- \frac{2}{N^2}  
\left( \ep \left[ \bI^\One, \Delta^\One \right] \right)  \bigg|_{\ep \to 0}
\label{identif}
\ee
with $\Delta^\Two (\ep)$ given by the finite contribution
of $ \bI^\Two (\ep) + \Delta^\One(\ep) \, \bI^\One (\ep)$
in \eqn{FItwo}.

Finally, by using eq.~(\ref{FIthree}) together with eq.~(\ref{Fthreedef}),
we obtain an expression for the three-loop Catani operator, namely
\bea
\bI^{\Three}(\ep)
&=&
- \third \left[ \bI^\One (\ep) \right]^3 
- \third \bI^\One (\ep) \bI^\Two (\ep)
- \frac{2}{3} \bI^\Two (\ep) \bI^\One (\ep) 
+ \bG^\Three (3 \ep)
\label{Ithree}
\\
&&+\frac{1}{3}\left\{ \left[  2 \bI^\Two (\ep) + \bI^\One (\ep) ^2, \Delta^\One (\ep) \right]
- \left[  \bI^\One (\ep) , \Delta^\One (\ep) \right] \Delta^\One (\ep)
+ \left[  \bI^\One (\ep) , \Delta^\Two (\ep) \right]\right\}
\nonumber
\eea
where all the commutator terms in the second line have 
a leading $\cO(1/\ep)$ divergence.
Since we do not have an explicit expression for $\bGam^\Three(\ep)$ in
$\bG^\Three(\ep)$,   we do not know the $\cO(1/\ep)$ contribution anyway so
we write
\be
\bI^{\Three}(\ep) =
- \third \left[ \bI^\One (\ep) \right]^3 
- \third \bI^\One (\ep) \bI^\Two (\ep)
- \frac{2}{3} \bI^\Two (\ep) \bI^\One (\ep) 
- \frac{11 \zeta_4  N^3}{(3\ep)^2} \bone 
+ \cO\left(\frac{1}{\ep}\right)  \,.
\label{Ithreeapprox}
\ee

In fact, we can check that the commutators in the second line of (\ref{Ithree}) don't have pieces proportional to the 
identity, thus we can absorb them inside the three-loop 
anomalous dimension matrix $\bGam^\Three$, as in \eqn{identif}. 
Thus we can also calculate the part of the 
$\cO(1/\ep)$ term proportional to the identity, 
\be
-\frac{\left(4\zeta_5+\frac{10}{3}\zeta_2\zeta_3\right)}{3\epsilon}
N^3
\bone
\ee
and we are left only with the unknown divergent piece $\bGam^\Three /\ep$.

If we substitute \eqn{Itwomatrix} into \eqn{Ithreeapprox}, 
we obtain
\bea
\bI^\Three(\ep)
&=
&\frac{1}{6}\left[\bI^\One(\ep)\right]^3
+\frac{N}{3}(\zeta_2+\epsilon \zeta_3)c(\epsilon)
\left[ \bI^\One(\ep) \bI^\One(2\ep) +2\bI^\One(2\ep)\bI^\One(\ep)\right]
\nonumber\\
&&-\frac{c(\ep)}{12\ep}
\left[ \bI^\One(\ep)\bhH^\Two+2\bhH^\Two\bI^\One(\ep)\right]
-\frac{11\zeta_4N^3}{9\ep^2}\bone+{\cal O}
\left(\frac{1}{\ep}\right)\,.
\label{Iexp}
\eea
Finally, we use \eqn{Ithreeapprox} to rewrite \eqn{threeloopCatani} as
\bea
\ket{A^\Three (\ep)} 
&=& 
  \frac{2}{3 N} \bI^\One(\ep)\ket{A^\Two (\ep)} 
+  \frac{1}{3 N^2} \bI^\Two(\ep)\ket{A^\One (\ep)} 
+   \frac{1}{3 N} \bI^\One(\ep)\ket{A^\Twof(\ep)} 
\nonumber\\
 &&+ \frac{2}{3 N^2} 
\left[ \bI^\Two (\ep) +\frac{1}{2} \bI^\One(\ep)^2 \right] 
\ket{A^\Onef(\ep)} 
- \frac{11 \zeta_4 }{9\ep^2} \ket{A^\Zero}
+ \cO\left(\frac{1}{\ep}\right) 
\eea
and then use \eqn{Itwomatrix} to obtain
\bea
\ket{A^\Three (\ep)} 
&=& 
  \frac{2}{3 N} \bI^\One(\ep)\ket{A^\Two (\ep)} 
+  \frac{1}{3 N^2} \bI^\Two(\ep)\ket{A^\One (\ep)} 
\label{athreeloop}\\
&&-   \frac{2}{3 \ep^2} \ket{A^\Twof(\ep)} 
+ \frac{\zeta_2}{3 \ep^2} \ket{A^\Onef(\ep)} 
- \frac{11 \zeta_4 }{9\ep^2} \ket{A^\Zero}
+ \cO\left(\frac{1}{\ep}\right) 
\nonumber
\eea
which will be useful in the following section.

\section{$1/N$ expansion of the IR divergences} 
\setcounter{equation}{0}
\label{secexpansion}

In this section, we will use the results of the previous section
to expand the IR-divergent contributions of the four-gluon
amplitude in powers of $1/N$.

First we re-express the vector of amplitudes (\ref{ketEll}) as
\be
\ket{A^\Ell}=  \left( \begin{array}{c} 
    \ket{A^\EllLC} + \frac{1}{N^2}\ket{A^\EllSC} + \cdots \\[3mm]
	\frac{1}{N} \ket{A^\EllDT} + 
	\frac{1}{N^3} \ket{A^\EllSDT} + \cdots 
\end{array} \right)
\label{block}
\ee
where 
\be
\ket{A^\Elleven}=\pmatrix{
A^\Elleven_\ii{1}\cr
A^\Elleven_\ii{2}\cr
A^\Elleven_\ii{3}\cr
A^\Elleven_\ii{4}\cr
A^\Elleven_\ii{5}\cr
A^\Elleven_\ii{6} \cr}
\qquad{\rm and}\qquad
\ket{A^\Ellodd}=\pmatrix{
A^\Ellodd_\ii{7}\cr
A^\Ellodd_\ii{8}\cr 
A^\Ellodd_\ii{9}}
\label{sixthree}
\ee
We recall that the leading-color amplitude $A^\EllLC$ is proportional to $N^L$ 
in the full amplitude $\cA_{4-{\rm gluon}}$  because
it is multiplied by $a^L \sim N^L$.
The subleading-color contributions $A^\Ellk$ are proportional 
to $N^{L-k}$ in $\cA_{4-{\rm gluon}}$, 
with  the most-subleading contribution $A^\EllL$
being the $N$-independent piece of the amplitude.

In the basis (\ref{block}) and (\ref{sixthree}), 
the operator $\bI^\One(\ep)$,
defined in \eqn{Ionedef},
has the form \cite{Glover:2001af}
\be
\bI^\One(\ep)  
= - \frac{1}{\ep^2}
\left( \begin{array}{cc}
N \ia & \ib \\
\ic  & N \id 
\end{array} \right)
\label{Ionematrix}
\ee
where 
\bea
\ia & = &
\left( {
\begin{array}{cccccc}
\tS+\tT & 0 & 0 & 0 & 0 & 0 \\
0 & \tS+\tU & 0 & 0 & 0 & 0 \\
0 & 0 & \tT+\tU & 0 & 0 & 0 \\
0 & 0 & 0 & \tT+\tU & 0 & 0\\
0 & 0 & 0 & 0 & \tS+\tU & 0\\
0 & 0 & 0 & 0 & 0 & \tS+\tT
\end{array} } \right), \qquad
\ib = 
\left( {
\begin{array}{ccc}
\tT-\tU & 0 & \tS-\tU \\
\tU-\tT & \tS-\tT & 0 \\
0 & \tT-\tS & \tU-\tS \\
0 & \tT-\tS & \tU-\tS \\
\tU-\tT & \tS-\tT & 0 \\
\tT-\tU & 0 & \tS-\tU \\
\end{array}
}\right)
\nonumber\\[4mm]
\ic & = &
\left( {
\begin{array}{cccccc}
\tS-\tU & \tS-\tT & 0 & 0 & \tS-\tT & \tS-\tU \\
0 & \tU-\tT & \tU-\tS & \tU-\tS & \tU-\tT& 0 \\
\tT-\tU & 0 & \tT-\tS & \tT-\tS & 0 & \tT-\tU \\
\end{array}
}\right), \qquad
\id = 
\left( {
\begin{array}{ccc}
2\tS & 0 & 0 \\
0 & 2\tU & 0 \\
0 & 0 & 2\tT
\end{array}
} \right)
\label{Ionesubmatrices}
\eea
with
\be
\tS = \left(-\frac{\mu^2}{s}\right)^\ep ,\qquad\qquad 
\tT = \left(-\frac{\mu^2}{t}\right)^\ep , \qquad\qquad 
\tU = \left(-\frac{\mu^2}{u}\right)^\ep  \,.
\ee

Using this together with \eqn{Hhatdef}, we can compute
\be
\bhH^\Two 
= \LOG \left( \begin{array}{cc}
\ha & N \hb \\
N \hc  & \hd 
\end{array} \right)
\label{Hhatmatrix}
\ee
with 
\bea
\label{Hsubmatrices}
\ha & = &
\left( {
\begin{array}{cccccc}
0 & 1 &-1 &-1 & 1 & 0 \\
-1& 0 & 1 & 1 & 0 &-1 \\
1 &-1 & 0 & 0 &-1 & 1 \\
1 &-1 & 0 & 0 &-1 & 1 \\
-1& 0 & 1 & 1 & 0 &-1 \\
0 & 1 &-1 &-1 & 1 & 0 \\
\end{array} } \right), \qquad
\hb = 
\left( {
\begin{array}{ccc}
1 & 0 &-1 \\
-1& 1 & 0 \\
0 &-1 & 1 \\
0 &-1 & 1 \\
-1& 1 & 0 \\
1 & 0 &-1 \\
\end{array}
}\right),
\nonumber\\[4mm]
\hc & = &
\left( {
\begin{array}{cccccc}
-1& 1 & 0 & 0 & 1 &-1 \\
0 &-1 & 1 & 1 &-1& 0 \\
1 & 0 &-1 &-1 & 0 & 1 \\
\end{array}
}\right). 
\eea

The leading-color amplitude $\ket{A^\EllLC}$
has poles up to $\cO(1/\ep^{2L})$, 
but from these expressions, we can see that 
each additional power of $1/N$ in the amplitude reduces
the power of the leading pole in $\epsilon$ by one,  
so that subleading-color amplitude $\ket{A^\Ellk}$
only has poles up to $\cO(1/\ep^{2L-k})$.
The diagonal (leading in $N$) elements of \eqn{Ionematrix}
have leading power $1/\ep^2$, whereas
the off-diagonal (subleading in $N$) elements 
have leading power $1/\ep$, 
since $\ib$ and $\ic$
have expansions that start at $\cO(\ep)$.
Since the leading divergences of $\bI^\Two$ and $\bI^\Three$ 
are given by 
$ -\half \left[ \bI^\One \right]^2$ and 
$ \frac{1}{6} \left[ \bI^\One \right]^3$ 
respectively,
one can use induction on 
eqs.~(\ref{oneloopCatani})--(\ref{threeloopCatani}) 
to show that the leading pole of $\ket{A^\Ellk}$ is $\cO(1/\ep^{2L-k})$.
We will see this explicitly in the following subsections.

\subsection{One-loop divergences}

We now substitute \eqns{block}{Ionematrix} into \eqn{oneloopCatani}
to obtain equations for the leading- and subleading-color  
one-loop amplitudes $A^\OneLC$ and $A^\OneDT$.
The leading-color amplitude satisfies
\be
\ket{A^\OneLC (\ep)} 
=
- \frac{1}{\ep^2} \ia \ket{A^\Zero} 
+ \ket{A^\OnefLC (\ep)}  \,.
\label{oneLC}
\ee
This equation is diagonal, so we focus on the first component 
\be
A_\ii{1}^\OneLC (\ep)
= - \frac{ 1}{\ep^2} (\tS + \tT ) A_\ii{1}^\Zero (\ep) 
+ A_\ii{1}^\OnefLC (\ep)
\label{oneLCone}
\ee
with the other components given by permutations of $s$, $t$, and $u$.
The finite contribution $ A_\ii{1}^\OnefLC (\ep) $ is not
specified by the IR analysis, 
but may be obtained by evaluating the exact expression 
(e.g., as in ref.~\cite{Bern:2005iz})
for the amplitude (\ref{Monedef})
\bea
A^\OneLC_\ii{1} &=&  M^\One (s,t) \, A^\Zero_\ii{1}
\nonumber\\
M^\One (s,t) &=& \, - \, \frac{ (\tS + \tT ) }{\ep^2}  
+ \half \log^2 \left( s\over t\right) + {2 \pi^2\over 3}
+ \cO(\ep) \,.
\label{Mone}
\eea

The equation for the subleading-color one-loop amplitude
\be 
\ket{ A^\OneDT (\ep)} 
=
- \frac{1}{\ep^2} \ic \ket{A^\Zero} + \ket{A^\OnefDT (\ep)}
\label{oneDT}
\ee
is consistent with the one-loop U(1) decoupling relation 
(\ref{oneloopdecouple}), but the latter (exact) relation
also allows us to evaluate the finite contribution
\be
A_\ii{7}^\OneDT (\ep)
=A_\ii{8}^\OneDT (\ep)
=A_\ii{9}^\OneDT (\ep)
=
\Atree \left[ \frac{(s\,\tS+ t\,\tT + u\,\tU)}{\ep^2} 
+ \half \left(  s X^2 + t Y^2 + u Z^2 \right) 
+\cO(\ep)
 \right]
\label{oneDTcomp}
\ee
where we define
\be
X = \log \left(t \over u\right), \qquad
Y = \log \left(u \over s\right), \qquad
Z = \log \left(s \over t\right).
\ee
We now expand $\tS$, $\tT$, and $\tU$ in $\ep$ 
and re-express 
\be
\log (-s/\mu^2) = \log (-u/\mu^2) -Y,
\qquad
\log (-t/\mu^2) = \log (-u/\mu^2) +X,
\qquad 
Z=-X-Y 
\ee
to obtain
\be
\ket{ A^\OneDT (\ep)} 
=
\Atree \left[ 
\left( \mu^2 \over -u \right)^\ep
\frac{(s Y - t X)}{\ep}  
- (s+t) XY 
 \right]  \ones
~+~\cO(\ep)
\label{oneDTexpand}
\ee
where $(1,1,1)^T$ 
indicates the $[7]$, $[8]$, and $[9]$ components
of $A^\OneDT$.
One can see that, while the leading-color
one-loop amplitude (\ref{oneLCone}) 
has an $\cO(1/\ep^2)$ leading divergence, 
the subleading-color amplitude (\ref{oneDTexpand}) 
has only an $\cO(1/\ep)$ pole.

\subsection{Two-loop divergences}

We derive expressions for the leading- and subleading-color
two-loop amplitudes 
$A^\TwoLC$, $A^\TwoDT$,  and $A^\TwoSC$
by substituting eqs.~(\ref{block}), (\ref{Ionematrix}), and (\ref{Hhatmatrix})
into \eqn{twoloopCataniagain}.

The IR behavior of the leading-color amplitude was utilized in 
ref.~\cite{Anastasiou:2003kj}
to motivate the ABDK relation between one- and two-loop
amplitudes.
To see this, observe that the leading-color amplitude satisfies
\bea
\ket{A^\TwoLC (\ep)} 
&=&
- \frac{1}{2 \ep^2} \ia 
\Biggl[ \ket{A^\OneLC (\ep)} + \ket{A^\OnefLC(\ep)}  \Biggr]
\nonumber
\\
&&
- (\zeta_2  + \ep \zeta_3) \cep 
\Biggl[ \ket{A^\OneLC (2 \ep)} - \ket{A^\OnefLC (2 \ep)}\Biggr]
+ \ket{A^\TwofLC (\ep)}
\label{twoLC}
\eea
the first component of which reads
\be
\label{twoLCone}
A_\ii{1}^\TwoLC (\ep) 
=
- \frac{ (\tS + \tT ) }{2 \ep^2}  
\left[ A_\ii{1}^\OneLC (\ep) + A_\ii{1}^\OnefLC(\ep)  \right]
- (\zeta_2  + \ep \zeta_3) \cep 
\left[ A_\ii{1}^\OneLC (2 \ep) - A_\ii{1}^\OnefLC (2 \ep)\right]
+ A_\ii{1}^\TwofLC (\ep) \,.
\ee
Using \eqns{twoloopLC}{Mone}, 
we may rewrite \eqn{twoLCone}  as
\be
M^\Two (\ep) 
\!=\!
\half 
\!\left[ M^\One (\ep)\! -\! M^\Onef (\ep)\right]
\!\left[ M^\One (\ep)\! + \!M^\Onef (\ep)\right]
- (\zeta_2  + \ep \zeta_3) \cep 
\left[ M^\One (2\ep) \!-\! M^\Onef (2\ep) \right] 
+ M^\Twof (\ep) 
\ee
where $ M^\Ellf(\ep)  = A^\EllfLC_\ii{1} (\ep) /  A^\Zero_\ii{1}$
and we have suppressed the $s$, $t$ dependence of $M^\Ell$.
Retaining only the divergent pieces, we get
\be
M^\Two (\ep) 
=
\half \left[ M^\One (\ep)\right]^2 
- (\zeta_2  + \ep \zeta_3)  M^\One (2\ep)  + \cO(\ep^0)\,.
\label{preABDK}
\ee
Of course, the Catani equation (\ref{twoloopCatani}) does not yield 
any information about the $\cO(\ep^0)$ piece, 
but 
Anastasiou et al. 
showed, using the exact 
one- and two-loop results (\ref{Monedef}) and (\ref{Mtwodef}),
that it is actually a constant 
(independent of the kinematic variables $s$ and $t$), 
yielding \cite{Anastasiou:2003kj} 
\be
M^\Two (\ep) 
=
\half \left[ M^\One (\ep)\right]^2 
- (\zeta_2  + \ep \zeta_3 + \ep^2 \zeta_4)  M^\One (2\ep)  - \frac{\pi^4}{72}
+ \cO(\ep)\,.
\label{ABDK}
\ee
The leading divergence of $M^\Two(\ep)$ is $\cO(1/\ep^4)$ as expected.

Next, the equation for the two-loop double-trace amplitude is
\bea
\ket{A^\TwoDT (\ep)} 
&=&
- \frac{1}{2\ep^2} \ic 
\Biggl[ \ket{A^\OneLC (\ep)} + \ket{A^\OnefLC (\ep)}  \Biggr]
- \frac{1}{2\ep^2} \id 
\Biggl[ \ket{A^\OneDT (\ep)} + \ket{A^\OnefDT (\ep)}  \Biggr]
\label{twoDT}\\
&&\hspace{-5mm} 
- (\zeta_2  + \ep \zeta_3) \cep 
\Biggl[ \ket{A^\OneDT (2 \ep)} - \ket{A^\OnefDT (2 \ep)}\Biggr]
+ \frac{\cep}{4 \ep} \LOG \hc \ket{A^\Zero}
+ \ket{A^\TwofDT (\ep)}
\nonumber\\
&=&
\Atree 
\frac{(-2)(s Y - t X)}{\ep^3}  
 \ones
~+~ \cO\left(1 \over \ep^2\right)
\eea
whose leading pole is $\cO(1/\ep^3)$.

Finally, the $N$-independent single-trace amplitude satisfies
\bea
\ket{A^\TwoSC (\ep)} 
&=& 
- \frac{1}{2 \ep^2} \ib 
\Biggl[ \ket{A^\OneDT (\ep)} + \ket{A^\OnefDT (\ep)} \Biggr]
+ \frac{\cep}{4 \ep} \LOG \ha \ket{A^\Zero}
+ \ket{A^\TwofSC (\ep)} \,.
\label{twoSC}
\eea
Using \eqns{Ionesubmatrices}{Hhatmatrix} and 
the first component of \eqn{twoSC} one obtains
\be
A_\ii{1}^\TwoSC (\ep)
= 
- \frac{(\tS + \tT - 2 \tU)}{2 \ep^2} 
\Biggl[ A_\ii{7}^\OneDT (\ep) + A_\ii{7}^\OnefDT (\ep) \Biggr]
+ \frac{\cep}{2 \ep} \LOG 
\Biggl[ A_\ii{2}^\Zero (\ep) - A_\ii{3}^\Zero (\ep) \Biggr]
+ A_\ii{1}^\TwofSC (\ep) \,.
\ee
Next we use \eqns{tree}{oneDTcomp} and expand in $\ep$ to obtain
the unexpectedly simple result
(due to cancellations between the $\left[ \bI^\One(\ep) \right]^2$
and
$\bhH^\Two$ terms in \eqn{Itwomatrix})
\be
A_\ii{1}^\TwoSC (\ep)
= 
\Atree \frac{X-Y}{\ep} \left[ 
\left( \mu^2 \over -u \right)^{2\ep}
\frac{s Y - t X}{2\ep}  
- (s+t) XY \right] + \cO(\ep^0)
\ee
which has an $\cO(1/\ep^2)$ leading divergence.
Comparing this with \eqn{oneDTexpand}, 
one obtains the following relation between 
the $N$-independent one- and two-loop amplitudes
\be
\ket{A^\TwoSC (\ep)} 
=
\frac{1}{\ep}
\pmatrix{X-Y \cr
         Z-X\cr 
         Y-Z\cr
         Y-Z\cr
         Z-X\cr 
	X-Y \cr}
A^{\OneDT}_{[7]}(2\ep) +
{\cal O}(\ep^0) \,.
\label{twoloopmat}
\ee
This relation manifestly obeys \eqn{sumSCvanishes}.
We have also verified \eqn{twoloopmat} 
using the exact two-loop amplitude (\ref{twoloopSC}).
Unlike the case of the ABDK relation (\ref{ABDK}) between 
leading-color one- and two-loop amplitudes,
however, the $\cO(\ep^0)$ term in \eqn{twoloopmat} is not
a simple constant, 
but rather a complicated linear combination of polylogarithms.

Finally, one can check that the equations
(\ref{twoLC}), (\ref{twoDT}), and (\ref{twoSC})
are consistent with the group theory relations (\ref{STDT}). 

\subsection{Three-loop divergences}

We derive expressions for the leading- and subleading-color
three-loop amplitudes 
$A^\ThreeLC$, $A^\ThreeDT$,  $A^\ThreeSC$, and $A^\ThreeSDT$
by substituting 
$\bI^\Two$ from (\ref{Itwomatrix}), 
$\bI^\One$ from (\ref{Ionematrix}),
and $\bhH^\Two$ from (\ref{Hhatmatrix})
into \eqn{athreeloop}.

The leading-color three-loop amplitude obeys
\bea
\ket{A^\ThreeLC (\ep)} 
&=&
- \frac{2}{3 \ep^2} \ia \ket{A^\TwoLC (\ep)} 
- \frac{1}{6 \ep^4} \ia^2 \ket{A^\OneLC (\ep)} 
+ \frac{(\zeta_2 + \ep \zeta_3) \cep}{12 \ep^2} \iatwo \ket{A^\OneLC (\ep)} 
\nonumber\\
&&
- \frac{2}{3\ep^2} \ket{A^\TwofLC (\ep)} 
+ \frac{\zeta_2}{3\ep^2} \ket{A^\OnefLC (\ep)} 
- \frac{11\zeta_4}{9\ep^4} \ket{A^\Zero}
+ \cO \left( \frac{1}{\ep} \right) \,.
\label{threeLC}
\eea
This equation may be shown to imply that 
\be
M^\Three (\ep) 
= M^\One (\ep) M^\Two(\ep) -
{1\over 3}\left[ M^\One (\ep)\right]^3 
- \left(11 \zeta_4\over 9 \ep^2\right) + \cO \left( 1 \over \ep\right)
\label{BDSthree}
\ee
which is consistent with eq.~(4.4) of ref.~\cite{Bern:2005iz},
though of course not as strong, since \eqn{BDSthree} was derived
from the IR behavior whereas the result of ref.~\cite{Bern:2005iz}
was derived by evaluating the exact three-loop amplitude.

The double-trace amplitude proportional to $N^2$ satisfies 
\bea
\ket{A^\ThreeDT (\ep)} 
&=&
- \frac{2}{3 \ep^2} \id \ket{A^\TwoDT (\ep)} 
- \frac{2}{3 \ep^2} \ic \ket{A^\TwoLC (\ep)} 
- \frac{1}{6 \ep^4} \id^2 \ket{A^\OneDT (\ep)} 
\nonumber\\ &&
- \frac{1}{6 \ep^4} \left( \ic \ia + \id \ic \right) \ket{A^\OneLC (\ep)} 
+ \frac{(\zeta_2 + \ep \zeta_3) \cep}{12 \ep^2} \idtwo \ket{A^\OneDT (\ep)} 
\nonumber\\ &&
+ \frac{(\zeta_2 + \ep \zeta_3) \cep}{12 \ep^2} \ictwo \ket{A^\OneLC (\ep)} 
+ \frac{\cep\LOG}{12 \ep} \hc \ket{A^\OneLC (\ep)} 
- \frac{2}{3\ep^2} \ket{A^\TwofDT (\ep)} 
\nonumber\\ &&
+ \frac{\zeta_2}{3\ep^2} \ket{A^\OnefDT (\ep)} 
+ \cO \left( \frac{1}{\ep} \right) 
\nonumber\\
&=&
\Atree 
\frac{2 (s Y - t X )}{\ep^5}   \ones
 +\cO\left(1 \over \ep^4\right) \,.
\label{threeDT}
\eea
The subleading-color single-trace amplitude satisfies 
\bea
\ket{A^\ThreeSC (\ep)} 
&=&
- \frac{2}{3 \ep^2} \ia \ket{A^\TwoSC (\ep)} 
- \frac{2}{3 \ep^2} \ib \ket{A^\TwoDT (\ep)} 
- \frac{1}{6 \ep^4} \left( \ia \ib + \ib \id \right) \ket{A^\OneDT (\ep)} 
\nonumber\\ &&
- \frac{1}{6 \ep^4} \ib \ic \ket{A^\OneLC (\ep)} 
+ \frac{(\zeta_2 + \ep \zeta_3) \cep}{12 \ep^2} \ibtwo \ket{A^\OneDT (\ep)} 
+ \frac{\cep\LOG}{12 \ep} \hb \ket{A^\OneDT (\ep)} 
\nonumber\\ &&
+ \frac{\cep\LOG}{12 \ep} \ha \ket{A^\OneLC (\ep)} 
- \frac{2}{3\ep^2} \ket{A^\TwofSC (\ep)} 
+ \cO \left( \frac{1}{\ep} \right) 
\nonumber\\
&=&
\Atree 
\frac{(-1)(  s Y - t X)}{\ep^4}  
\pmatrix{X-Y \cr
         Z-X\cr 
         Y-Z\cr
         Y-Z\cr
         Z-X\cr 
	 X-Y\cr}
 +\cO\left(1 \over \ep^3\right)\,.
\label{threeSC}
\eea
Finally, the $N$-independent double-trace amplitude obeys
\bea
\ket{A^\ThreeSDT (\ep)} 
&=&
- \frac{2}{3 \ep^2} \ic \ket{A^\TwoSC (\ep)} 
- \frac{1}{6 \ep^4} \ic \ib \ket{A^\OneDT (\ep)} 
+ \cO \left( \frac{1}{\ep} \right) 
\nonumber\\
&=& 
 \frac{1}{6 \ep^4} \ic \ib \left[ \ket{A^\OneDT (\ep)} 
                       + 2 \ket{A^\OnefDT (\ep)}  \right]
-\frac{\cep \LOG}{6 \ep^3} \ic \ha \ket{A^\Zero}
+ \cO \left( \frac{1}{\ep} \right) \,.
\label{athreethree}
\eea
We use 
\eqn{oneDTexpand}
together with  $L =- XY(X+Y)$ and
\be
\ic\ib \ones  =  
2 \left[  (S-T)^2 +  (T-U)^2 +  (U-S)^2  \right] \ones
\label{gammabeta}
\ee
and expand in $\ep$ to obtain the leading two 
terms in the Laurent expansion 
\be
\ket{A^\ThreeSDT (\ep)} 
=
\Atree 
{ X^2 + Y^2 + Z^2 \over \ep^2 } 
\left[ 
\left( \mu^2 \over -u \right)^{3\ep}
\frac{s Y - t X}{3\ep}  
- (s+t) XY 
\right]  \ones
+ \cO \left(1 \over \ep \right) \,.
\label{threeeps2}
\ee
This can be written concisely as
\be
\ket{A^\ThreeSDT (\ep)} 
=
{ X^2 + Y^2 + Z^2 \over \ep^2 } 
\ket{A^\OneDT (3\ep)} 
+ \cO \left(1 \over \ep \right)\,.
\label{threeeps}
\ee
One can see from all these expressions that 
$\ket{A^\Threek}$ has a leading pole of $\cO(1/\ep^{6-k})$, as expected.

\subsection{Higher-loop divergences}

Equations (\ref{Itwomatrix}) and (\ref{Iexp}) suggest that
the most-divergent contribution of the $L$-loop amplitude is given by
\be
\ket{A^\Ell (\ep)} 
= {1 \over L!} 
\left[ {\bI^\One(\ep)  \over N} \right]^L  \ket{A^\Zero} +
 \cdots
\label{conjec}
\ee
which of course can be summed to give 
\be
\ket{A(\ep)} 
= \exp \left[  \bI^\One(\ep) \over N \right]  \ket{A^\Zero} +
 \cdots
\label{expconj}
\ee
Equation (\ref{conjec})
is certainly valid for the 
leading-color contribution $\ket{A^\EllLC}$, as it implies
\be
M^\Ell = {1 \over L!} \left[ M^\One \right]^L + \cdots
\ee
the leading-term of the BDS relation \cite{Bern:2005iz}.
Our calculations in previous subsections,
however, 
show that \eqn{conjec} also correctly gives the 
most-divergent $\cO(1/\ep^{2L-k})$ contribution 
of the subleading-amplitudes $\ket{A^\Ellk}$,
at least for $L \le 3$.
We expect this pattern to continue to higher loops. 
For example, the leading divergence of  the $N$-independent
amplitude $A^\EllL$ should be given by 
\be
 {1 \over N^L  L!} 
\left[ \bI^\One(\ep)  \Big|_{\rm N-indep} \right]^L
= 
\frac{(-1)^L}{N^L\, L!\,\ep^{2L}}\left( \begin{array}{cc}
 0 & \ib \\
\ic  &  0 
\end{array} \right)^L+{\cal O}\left(\frac{1}{\ep^{L-1}}\right)
\label{generaliz}
\ee
where $\gamma_\ep$ and $\beta_\ep$ are of $\cO(\ep)$, 
so that the leading divergence is of $\cO(1/\ep^L)$.

We now treat the cases $L=2k+1$ and $L=2k+2$ separately. 
For $L=2k+1$, 
using \eqns{oneDTexpand}{gammabeta},
\eqn{conjec} implies that the leading divergence
of $A^{(2k+1,2k+1)}(\ep)$ is given by
 \bea
 \ket{A^{(2k+1,2k+1)}(\ep)}
 &=&
 \frac{1}{(2k+1)!\, \ep^{4k}}
 (\ic\ib )^k 
 \ket{A^\OneDT (\ep)} 
 +\cO \left(1 \over \ep^{2k} \right)
\label{LLIRdiv}
 \\
 &=&
\frac{2^k}{(2k+1)!}
 \Atree 
\left[\frac{X^2+Y^2+Z^2}{\ep^2}\right]^k
\left[ \frac{s Y - t X}{\ep} \right]  
\ones
+  \cO \left(  1 \over \ep^{2k}  \right)
\nonumber
\eea
which can be formally summed to give
 \be
\sum_{k=0}^\infty \left(a\over N\right)^{2k+1} \ket{A^{(2k+1,2k+1)}(\ep)}
= \Atree 
\sinh \left(a  \sqrt{ 2 (X^2 + Y^2+ Z^2) } \over N \ep \right)
\left[ \frac{s Y - t X}{\sqrt{2 (X^2+ Y^2 + Z^2) } } \right]  
\ones \,.
\ee
For $2k+2$, there is one more $\bI^\One$ matrix acting, 
and we get
\be
 \ket{A^{(2k+2,2k+2)}(\ep)}
=
-\frac{1}{(2k+2)!\, \ep^{4k+2}}
\ib 
(\ic \ib)^k 
 \ket{A^\OneDT (\ep)} 
+\cO \left(\frac{1}{\ep^{2k+1}}\right)
\ee
and using 
\be
\ib\ones
=\epsilon\pmatrix{Y-X\cr X-Z\cr Z-Y\cr Z-Y\cr X-Z\cr Y-X}+\cO (\epsilon^2)
\ee
we find that the leading divergence of $A^{(2k+2,2k+2)}(\ep)$ is given by
\bea
 \ket{A^{(2k+2,2k+2)}(\ep)}
&=&
\frac{2^k}{(2k+2)!}
\Atree 
\frac{1}{\ep}
\left[\frac{X^2+Y^2+Z^2}{\epsilon^2}\right]^k
\left[ \frac{s Y - t X}{\ep} \right]  
\pmatrix{ X-Y\cr Z-X\cr Y-Z \cr Y-Z \cr Z-X\cr X-Y}
+\cO \left(\frac{1}{\epsilon^{2k+1}}\right)
\nonumber\\
\label{LLIRdiv2}
\eea
which can also be formally summed to give
 \bea
\sum_{k=0}^\infty \left(a\over N\right)^{2k+2} \ket{A^{(2k+2,2k+2)}(\ep)}
&=& \Atree 
\left[ s Y - t X \over 2 (X^2+ Y^2 + Z^2)  \right]    \times
\\
&& \qquad
\left[ \cosh \left(a  \sqrt{ 2 (X^2 + Y^2+ Z^2) } \over N \ep \right) - 1\right]
\pmatrix{ X-Y\cr Z-X\cr Y-Z \cr Y-Z \cr Z-X\cr X-Y}\,.
\nonumber
\eea
Based on the forms of \eqns{twoloopmat}{threeeps},
we make the stronger conjecture
that the first two terms in the Laurent expansions of the
$N$-independent amplitudes are given by 
\bea
 \ket{A^{(2k+1,2k+1)}(\ep)}
&=&
\frac{ 2^k} { (2k)! } \left[ X^2 + Y^2 + Z^2 \over \ep^2  \right]^k
A^{\OneDT}_{[7]}((2k+1)\ep) 
\ones 
+ \cO \left(1 \over \ep^{2k-1} \right)
\label{Lodd}
\\
 \ket{A^{(2k+2,2k+2)}(\ep)}
&=&
\frac{2^k}{(2k+1)!}
\frac{1}{\ep}
\left[ X^2 + Y^2 + Z^2 \over \ep^2  \right]^k
A^{\OneDT}_{[7]}((2k+2)\ep) 
\pmatrix{X-Y \cr Z-X\cr Y-Z\cr Y-Z\cr Z-X\cr X-Y \cr}
+ {\cal O}\left( 1 \over \ep^{2k} \right)
\nonumber
\\
\label{Leven}
\eea
but we have not tried to verify these.
These equations of course would imply that
\bea
\sum_{i=7}^9 A^{(2k+1,2k+1)}_\ii{i} (\ep)
&=&
3 \frac{ 2^k} { (2k)! } \left[ X^2 + Y^2 + Z^2 \over \ep^2  \right]^k
A^{\OneDT}_{[7]}((2k+1)\ep) 
+ \cO \left(1 \over \ep^{2k-1} \right)
\\
\sum_{i=1}^6 A^{(2k+2,2k+2)}_\ii{i} (\ep)
&=& {\cal O}\left( 1 \over \ep^{2k} \right)\,.
\eea

The exponentiation property (\ref{expconj}),
which implies that the leading $L$-loop divergence 
of the $N$-independent amplitudes is $\cO(1/\ep^L)$, 
reminds us of similar behavior 
in ${\cal N}=8$ supergravity \cite{Naculich:2008ew,Brandhuber:2008tf},
so it is natural to try to relate 
the $N$-independent $\cN=4$ SYM amplitudes 
to ${\cal N}=8$ supergravity amplitudes.

\section{$\cN=4$ SYM / $\cN=8$ supergravity connection}
\setcounter{equation}{0}
\label{secSG}

In this section, we demonstrate the existence 
of some relations between 
$\cN=4$ SYM amplitudes  and
$\cN=8$ supergravity amplitudes at the one- and two-loop levels. 
The $L$-loop $N$-independent SYM amplitude $A^\EllL$ is related to
the $L$-loop supergravity amplitude, 
as both have $\cO(1/\ep^L)$ leading IR divergences. 
Other subleading-color SYM amplitudes $A^\Ellk$ 
have $\cO(1/\ep^{2L-k})$ leading IR divergences, 
and consequently satisfy relations involving 
lower-loop supergravity amplitudes.

In this section we use the notation\footnote{The normalization
of $\Asy^\Ellodd (s,t)$ is arbitrary.  
We have chosen one that is most natural in the context of
the SYM/supergravity relations presented in this section.}
\be
\Asy^\Elleven (s,t) = a^L A^\Elleven_{[1]}, \qquad
\Asy^\Ellodd (s,t) = - {a^L \over \sqrt 2} A^\Ellodd_{[8]}
\ee
noting that the other components $A^\Ellk_{[i]}$ 
are obtained by permutations of  $s$, $t$, and $u$.
However, we omit the argument $(s,t)$ for functions that
are completely symmetric under permutations of $s$, $t$, and $u$.
We also define
\be
\Msy^\Ellk (s,t)=  {\Asy^\Ellk (s,t)\over \Asy^\Zero(s,t)} \,.
\ee
Note that the coupling constant $a^L$ is now included in the definition
of $\Msy^\Ellk (s,t)$ 
(as it is for the supergravity amplitudes in ref.~\cite{Naculich:2008ew})   
in order to make the supergravity--nonplanar SYM relations more 
transparent.
This differs from $M^\Ell$ defined in previous sections,
so that $\Msy^\EllLC = a^L M^\Ell$.

\subsection{One- and two-loop relations}

Recall that the one-loop $N$-independent SYM four-gluon amplitude is
given by (\ref{oneloopnp})
\be
\Asy^\OneDT = -\frac{a}{\sqrt2} A^\OneDT_\ii{8} =
- 2 \sqrt2 i K 
\left[\frac{g^2 N}{8\pi^2}\left(4\pi \e^{-\gamma}\right)^\ep\right] 
\left[ I_4^\One (s,t)+ I_4^\One (t,u)+ I_4^\One (u,s) \right]\,.
\ee
The one-loop supergravity four-graviton amplitude\footnote{after 
stripping off a factor of $(\kappa/2)^2$ for a four-point amplitude}
may be expressed as \cite{Green:1982sw,Bern:1998ug}
\be
\Asgone
=8iK^2 
\left[\frac{(\kappa/2)^2}{8\pi^2}\left(4\pi \e^{-\gamma}\right)^\ep\right] 
 \left[ I_4^\One (s,t)+ I_4^\One (t,u)+ I_4^\One (u,s) \right]\,.
\ee
The supergravity amplitude is proportional to $K^2$ rather than $K$ 
due to the KLT relations  \cite{Kawai:1985xq}
(a manifestation of the relation ``closed string = (open string)$^2$").
Defining $\lSYM = g^2 N$ and $\lSG = (\kappa/2)^2$,
one observes that the one-loop SYM and supergravity amplitudes are related by
\be
\Asy^\OneDT =- \frac{1}{2 \sqrt2 K}\frac{\lSYM}{\lSG }\Asgone\,.
\label{oneloopArelation}
\ee
By factoring out the tree amplitudes in both the supergravity 
and SYM amplitudes
\bea
\Asgone &=&    \Asgzero \Msgone   =\left(16iK^2\over stu\right) \Msgone
\\
\Asy^\OneDT &=&  \Asy^\Zero(s,t)\Msy^\OneDT(s,t)  
= \left( - \frac{4iK}{st}\right) \Msy^\OneDT(s,t)
\eea
we can express \eqn{oneloopArelation} in the form
\be
\Msy^\OneDT(s,t)= \sqrt2 \, \frac{\lSYM}{\lSG  u}\Msgone\,.
\label{oneloopMrelation}
\ee
In other words, 
the ratio of the one-loop subleading-color SYM 
and the one-loop supergravity amplitudes 
(after factoring out the tree amplitudes) 
is simply proportional to the ratio of coupling constants, 
where we need to use the effective dimensionless coupling $\lSG u$
for supergravity because $\lSG$ is dimensionful. 

Finally, we rewrite \eqn{oneloopMrelation} in the manifestly
permutation-symmetric form
\be
\frac{1}{3}\left[(\lSG u)\Msy^\OneDT(s,t) + \rmcp \right]
=\sqrt2\lSYM\Msgone
\label{oneloopsugra}
\ee
(where $\rmcp$ denotes cyclic permutations of $s$, $t$, and $u$)
even though $u\Msy^\OneDT(s,t)$ is already symmetric under permutations.
A similar symmetrized relation can be written for the 
one-loop leading-color amplitude 
\be
(\lSG u)\Msy^\OneLC(s,t)+\rmcp 
=-\lSYM\Msgone
\label{onelooprelat}
\ee
obtained from the one-loop decoupling relation (\ref{oneloopdecouple}) 
together with \eqn{oneloopMrelation}. 

We now turn to {\it two loops}.
First, we exhibit some relations between SYM and 
supergravity amplitudes that hold only for the IR-divergent terms.
The easiest case to analyze is the
two-loop $N$-independent SYM amplitude $\Asy^\TwoSC (s,t)$, 
since from \eqn{twoloopmat} we have
\be
\Asy^\TwoSC (s,t) = a^2 A^\TwoSC_\ii{1} =
- \sqrt2 a  \frac{X-Y}{\ep} \Asy^{\OneDT}(2\ep) + {\cal O}(\ep^0)\,.
\label{twosc}
\ee
Using \eqn{oneloopArelation}, we can rewrite this as
\be
\Asy^\TwoSC (s,t) =
\frac{a}{2K}\frac{\lSYM}{\lSG }
\left( X-Y\over \ep\right)  
\Asgone(2\ep)+{\cal O}(\ep^0)
\ee
or equivalently
\be 
\Msy^\TwoSC (s,t)
=
-2 a \,\frac{\lSYM}{\lSG u}
\left(\frac{X-Y}{\ep} \right)
\Msgone(2\ep)+{\cal O}(\ep^0)
\label{IRSC}
\ee
thus obtaining a relation to the one-loop supergravity amplitude.

The IR-divergent part of $\Asy^\TwoDT (s,t)$ 
is also related to the one-loop supergravity amplitude, 
though in a rather more complicated way.
First, it can be related to the
one-loop subleading-color amplitude $\Asy^\OneDT (\ep)$
by the expression
\be
\Asy^\TwoDT (s,t) =
a \left( \mu^2 \over t \right)^{\ep} 
\left[  \left( - \, {2\over \ep^2} 
     +  {7 \pi^2 \over 12} \right) \Asy^\OneDT (\ep)
        - {1\over 2\ep^2} \left( \Asy^\OneDT \Big|_{\cO(\ep^0)}  \right) \right]
 - {3 a  X \over \ep} \Asy^\OneDT (2 \ep)
+ \cO ( \ep^0 )
\ee
obtained from an explicit evaluation of the IR-divergent part 
of the scalar integrals in \eqn{twoloopDT} in the physical region 
$t>0$,  $s,u<0$.
Then, by virtue of \eqn{oneloopArelation}, it can be related
to the one-loop supergravity amplitude by
\bea
\Msy^\TwoDT (s,t) &=&
\sqrt2  {a \lSYM \over \lSG u} 
\Bigg\{ \left( \mu^2 \over t \right)^{\ep} 
\left[  \left( - \, {2\over \ep^2} 
     +  {7 \pi^2 \over 12} \right) \Msgone (\ep)
        - {1\over 2\ep^2} \left( \Msgone  \Big|_{\cO(\ep^0)}  \right) \right]
\nonumber\\
&&\hspace{25mm} - {3 X \over \ep} \Msgone  (2 \ep) \Bigg\}
+ \cO ( \ep^0 )\,.
\label{IR21}
\eea

Next, we exhibit some two-loop relations 
to ${\cN}=8$ supergravity that include the finite terms. 
First, we consider the two-loop $N$-independent amplitude $\Msy^\TwoSC$. 
Multiplying \eqn{IRSC} by $u^2$ and summing over cyclic permutations, 
we can write 
\be
\frac{1}{3}
\left[(\lSG u)^2\Msy^\TwoSC  (s,t) + \rmcp \right]
=  \lSYM^2 \left[\Msgone (\ep) \right]^2  + \cO(\ep^0)\,.
\ee
Then, using the relation 
$\Msgtwo (\ep) =\half [\Msgone(\ep) ]^2+ \cO(\ep^0)$ 
between the one- and two-loop supergravity 
amplitudes \cite{Naculich:2008ew,Brandhuber:2008tf},
we can write this as  
\be
\frac{1}{3}
\left[(\lSG u)^2\Msy^\TwoSC(s,t) + \rmcp \right]
= 2\lSYM^2 \Msgtwo
\label{twoloopsugra}
\ee
where we omit the $\cO(\ep^0)$ because, 
in fact,  this  relation is exact (!),
as may be easily verified by using the
exact expression (\ref{twoloopSC}) for the $N$-independent 
SYM amplitude \cite{Bern:1997nh} 
\bea
\Msy^\TwoSC (s,t) 
&=& 
\frac{a^2 s t}{2} 
\Bigl[
s \left(I_4^\TwoP(s, t) + I_4^\TwoNP(s, t)
+ I_4^\TwoP(s, u)  + I_4^\TwoNP(s, u)\right)
\label{twoloopSG}\\ && \hspace{11mm}
+t \left(I_4^\TwoP(t, s) + I_4^\TwoNP(t, s)
+ I_4^\TwoP(t, u)  + I_4^\TwoNP(t, u)\right)
\nonumber\\ && \hspace{10mm}
-2u \left(I_4^\TwoP(u, s) + I_4^\TwoNP(u, s)
+ I_4^\TwoP(u,t)  + I_4^\TwoNP(u,t)\right)
\Bigr]  
\nonumber
\eea
and that for the two-loop supergravity amplitude \cite{Bern:1998ug}
\be
\Msgtwo 
= - \frac{s^3 t u }{4} 
\left[\frac{(\kappa/2)^2}{8\pi^2}
\left(4\pi \e^{-\gamma} \right)^\ep\right]^2
[I_4^{(2)P}(s,t)+I_4^{(2)NP}(s,t)
+I_4^{(2)P}(s,u)+I_4^{(2)NP}(s,u)]+ \rmcp
\ee

Finally we turn to the two-loop subleading-color amplitude $\Msy^\TwoDT$. 
The two-loop decoupling relation 
(\ref{twoloopdecouple}) can be rewritten as 
\be
-\sqrt2 \left[ u\Msy^\TwoDT (s,t)+ \rmcp \right]
=6 \left[ u\Msy^\TwoLC (s,t) + \rmcp \right]\,.
\label{rela}
\ee
Using the ABDK relation \cite{Anastasiou:2003kj}
\be
\Msy^\TwoLC (\ep) =
\frac{1}{2}\left[\Msy^\OneLC (\ep) \right]^2
+ a f^\Two(\ep)\Msy^\OneLC (2\ep)+\cO(\ep), \qquad
f^\Two(\ep) = - (\zeta_2  + \ep \zeta_3 + \ep^2 \zeta_4)
\ee
together with \eqn{onelooprelat},
we can rewrite \eqn{rela} as
\bea
\frac{1}{3}\left[(\lSG u)\Msy^\TwoDT(s,t)+\rmcp \right]
&=& - \frac{1}{\sqrt2} \left\{
(\lSG u)\left[\Msy^\OneLC(s,t)\right]^2+ \rmcp \right\}
\nonumber\\
&&
+ \sqrt2 \frac{\lSYM^2}{8\pi^2} \left(4\pi \e^{-\gamma}\right)^\ep
 f^{(2)}(\ep)\Msgone(2\ep) +\cO(\ep)\,.
\label{twoloopre}
\eea
Unlike the previous relation, 
however, \eqn{twoloopre} only holds through $\cO(\ep^0)$.
As mentioned above,
because of the fact that 
the leading IR divergence of $\Msy^\TwoDT$ is $\cO(1/\ep^3)$, 
we found a relation to the one-loop supergravity amplitude rather
than the two-loop one. 

In the next subsection, we attempt to generalize the exact relations
(\ref{oneloopsugra}) and (\ref{twoloopsugra}) to $L$ loops. 
We might try to generalize \eqn{twoloopre} as well, 
but it was based on the two-loop decoupling relation (\ref{twoloopdecouple}),
which does not hold beyond two loops.

\subsection{Looking for a general ansatz 
and a connection to the 't Hooft picture}

The relations (\ref{oneloopsugra}) and (\ref{twoloopsugra}) 
between $N$-independent SYM amplitudes and supergravity amplitudes at
one and two loops suggest the appealing generalization
\be
\frac{1}{3}
\left[(\lSG u)^L\Msy^\EllL(s,t) + \rmcp \right]
 \quad \stackrel{?}{=} \quad (\sqrt2\lSYM )^L \MsgEll 
\label{llloop}
\ee
which is exact for $L=0$, 1, and 2. 
It is tempting to hope that this relation holds for all $L$. 
Unfortunately, \eqn{llloop} fails starting at $L=3$,
even at leading order,  $\cO(1/\ep^L)$.

In sec.~4.3, we used the three-loop formula of 
Sterman and Tejeda-Yeomans 
to derive the first two terms in the Laurent expansion of 
$A^\ThreeSDT$ in \eqns{threeeps2}{threeeps}.
On the supergravity side, 
we expect \cite{Naculich:2008ew} that,
at least at leading IR order, we have an exponentiation formula, i.e. 
\be
\MsgEll 
= \frac{1}{L!} \left[\Msgone  \right]^L + \cO \left( 1 \over \ep^{L-1} \right)
= \frac{1}{L!} \left[- \lSG (s Y - tX) \over  8 \pi^2 \ep \right]^L 
+ \cO \left( 1 \over \ep^{L-1} \right)\,.
\label{sugrairdiv}
\ee
One can explicitly check that, if \eqn{sugrairdiv} is true for $L=3$,
then \eqn{llloop} is not satisfied at three loops.\footnote{The three-loop 
formula of Sterman and Tejeda-Yeomans was not 
derived by an explicit calculation, but its $1/\ep^3$ term, 
which gives the leading term in $A^\ThreeSDT$, is probably correct. 
Similarly, we did no explicit three-loop calculation 
for ${\cal N}=8$ supergravity, and the exponentiation conjecture 
for the leading IR divergence in ref.~\cite{Naculich:2008ew} was
based on the two-loop exponentiation formula and 
the fact that in the theories where this was studied, 
at least the leading IR divergences exponentiate. 
So it is possible, but very unlikely, that \eqn{llloop} 
holds beyond two loops.}

Assuming that \eqn{sugrairdiv} for the leading IR 
divergence of the supergravity amplitude is correct,  
and that the leading IR divergences of $A^\EllL$ 
conjectured in \eqns{LLIRdiv}{LLIRdiv2} are also correct, 
one can show that
the following relations hold:
\bea
&&\left[\lSG^2\frac{s^2+t^2+u^2}{3}\right]^k
\frac{1}{3}\left[(\lSG u) \Msy^{(2k+1,2k+1)} (s,t;\ep)+\rmcp\right]
\nonumber\\
&&\hspace{-1.6cm}
=\lSYM^{2k+1} \frac{2^{2k+1/2}}
{(2k+1)!}
\left[\Msgtwo (\ep) +\frac{1}{6} 
\left( \lSG \over 8 \pi^2 \right)^2 
\left(\frac{sX+tY+uZ}{\ep}\right)^2
\right]^k
\Msgone (\ep)+\cO  \left(\frac{1}{\ep^{2k}}\right)
\label{2k1}
\eea
for $L=2k+1$ and
\bea
&&
\left[\lSG^2 \frac{s^2+t^2+u^2}{3}\right]^k
\frac{1}{3}
\left[(\lSG u)^2 \Msy^{(2k+2,2k+2)} (s,t;\ep)+\rmcp \right]
\nonumber\\
&&\hspace{-1.9cm}
=\lSYM^{2k+2} \frac{2^{2k+2}}{(2k+2)!}
\left[\Msgtwo (\ep)+\frac{1}{6}
\left( \lSG \over 8 \pi^2 \right)^2 
\left(\frac{sX+tY+uZ}{\ep}\right)^2
\right]^k \Msgtwo (\ep)+\cO \left(\frac{1}{\ep^{2k+1}}\right)
\label{2k2}
\eea
for $L=2k+2$ (where $k=0,1,2,...$) 
instead of the result \eqn{llloop} 
without the correction to $\Msgtwo$ inside the square brackets, 
and with $[\frac{1}{3} (s^2+t^2+u^2)]^k$ replaced by $u^{2k}$.

An interesting fact is that either \eqn{llloop} 
(or \eqns{2k1}{2k2} without the extra term), 
and also the relation (\ref{twoloopre}),  
have a possible interpretation in terms of the 
't Hooft string picture of the $1/N$ expansion. 
Thus at least in the case of $L=1,2$,  \eqns{llloop}{twoloopre} 
still do, so one can hope that there is a
correct relation at higher $L$ yet to be determined.

't Hooft's idea was to construct string worldsheets 
out of Yang-Mills Feynman diagrams by drawing simplified diagrams 
(depending only on a particle's color structure),
with adjoint fields represented by double lines
and fundamental fields (quarks) by single lines.
In $\cN=4 $ SYM, all fields are in the adjoint representation,
therefore all 't Hooft diagrams are composed of double lines only.
In this picture,
an index line loop represents a color trace,\footnote{The
$g$- and $N$-dependence
of an $n$-gluon amplitude can be written suggestively as 
$g^n (g^2N)^I (g^2)^{2H+B-2}$, 
where $I$=number of index loops, $H$=number of handles,
and $B$=number of boundaries of the Feynman diagram=worldsheet.
Each index loop brings a factor of $N$, 
as well as a factor of $g^2$, and
$2H+B-2$ characterizes the topology of the surface. 
Thus an index loop is associated with the same coupling behaviour 
as an open string loop (splitting and rejoining). }
contributing a factor of $N$.

As we noted earlier, the leading-color amplitudes correspond to planar diagrams,
carrying a factor of $N^L$ at $L$ loops.
A subleading-color contribution down by $1/N$
(e.g., the leading term of the double-trace amplitude $A_{4;3}$)
comes from a diagram missing a line loop (= color trace),
which for the 't Hooft diagrams of $\cN=4$ SYM can
only come from twists of the external 
(on the outside boundary of the corresponding planar diagram)
double lines\footnote{
For example, the one-loop four-gluon diagram
contains three possible cases, 
with two twists adjacent to an external gluon $a_1$
giving $\Tr(T^{a_1}) \Tr( T^{a_2}T^{a_3} T^{a_4})$,
which vanishes for \SUN, 
two twists on opposite side of the box giving, for example,
$  \Tr(T^{a_1} T^{a_2}) \Tr(T^{a_3} T^{a_4})$,
and twists on all four double lines giving 
$  \Tr(T^{a_1} T^{a_3}) \Tr(T^{a_2} T^{a_4})$.}
in such a way that index loops of the external gluons become disconnected,
giving a diagram with the topology of a hole (annulus).
A subleading-color amplitude down by $1/N^2$
corresponds to a diagram which also has twists of 
internal (not belonging to the outside boundary of the planar diagram)
double lines 
giving a nonplanar diagram with a handle,
which does not modify the external color trace.
For higher-point functions, we could have 
multiple ($k+1$) trace amplitudes down by $1/N^k$,
coming from diagrams with the topology of a surface 
with $k$ holes (open string loops). 

't Hooft's proposal for a relation to string theory 
associated a set of Yang-Mills diagrams with a string worldsheet. 
(The original idea for the string to live in four flat dimensions 
never quite worked out in detail,
though for ${\cal N}=4$ SYM theory, 
the $AdS_5\times S^5$ string may be the correct construction.) 
Thus, a planar (leading-color) SYM diagram corresponds to a 
tree-level string worldsheet,
a $1/N$ subleading-color SYM diagram with the topology of a hole (annulus)
corresponds to an open-string one-loop worldsheet,
and a $1/N^2$ subleading-color SYM diagram with a handle topology 
corresponds to a closed-string one-loop worldsheet.

It is possible, 
and our two-loop relation between
 $\Asy^\EllL$ and $\AsgEll$ for $L=1,2$ seems to suggest this, 
that one can reduce each closed-string loop to two open-string 
loops (this relation is certainly valid for vacuum diagrams).
The open-string loop comes with a factor of 
$\gSYM^2=g_{\rm open}^2=g_s$, 
whereas  a closed-string diagram comes with a 
factor of $\gSYM^4=g_{\rm closed}^2=g_s^2$, 
so that one would have 
\be
{\rm one\; loop\;\;closed\;string\;=\;(one \;loop\;\;open\;string)^2}
\ee
reducing each $1/N$ factor in the SYM amplitudes to an open-string loop; 
thus $A^\Ellk$ corresponds to $k$ open-string loops.
But now, if we also adopt the rule
\be
{\rm one\;loop \;open\  't\;Hooft\;string\;(in\; 4d!) = \;
one\;loop \;in\;\; {\cN}=8\; supergravity\;in \;4d}
\label{rule}
\ee
then \eqns{twoloopre}{llloop} 
(or \eqns{2k1}{2k2} with no extra terms)
have a simple interpretation:  
$A^{(L,k)}$ corresponds to $k$ loops in the supergravity expansion.

Using the rule (\ref{rule}) for the $k=L$ relation (\ref{llloop}), 
for which the number of loops in SYM equals the 
number of loops in supergravity, 
there are no internal loops in the 't Hooft double line diagram, 
thus the diagram is defined exclusively by its handles and holes.
Then \eqn{llloop} just corresponds to replacing
$\Msy^\EllL$  with $\MsgEll$, 
$\sqrt{2}\lSYM$ with $\lSG u$ 
(the effective supergravity coupling constant) and, 
since supergravity amplitudes are permutation-symmetric 
whereas SYM amplitudes are not 
(the position of the twists breaks the 
symmetry of the Feynman diagram, 
even if nothing else does), 
averaging over cyclic permutations. 
If the correction term  were not present 
in the square brackets in \eqns{2k1}{2k2}, 
then we would just need to replace 
$\lSYM^{2k+1}$ with $(\lSG^2 (s^2+t^2+u^2)/3)^k \lSG u$ and 
$\lSYM^{2k+2}$ with $(\lSG^2 (s^2+t^2+u^2)/3)^k (\lSG u)^2$ instead. 
Since the correction terms are present, the interpretation is not clear.

It is also not clear how to derive the rule (\ref{rule}), 
or why such a rule should even be possible. 
It is reminiscent of AdS/CFT, 
but then it is not clear why we get supergravity in 4d 
and not some higher-dimensional space. 
If true, it could be a manifestation of a different kind of duality, 
relating weak coupling with weak coupling, 
as also advocated in ref.~\cite{ArkaniHamed:2008gz}.

On the other hand, perhaps the rule (\ref{rule}) is only an approximation. 
When compactifying string theory down to 4d, 
the supergravity modes could in principle get mixed up with 
other string modes, and the loop expansion of supergravity 
combined with other terms, 
so it is in principle possible that a modification of the rule (\ref{rule}) 
could account for the modified relations (\ref{2k1}) and (\ref{2k2}), 
and make them precise beyond leading order.

\section{Transcendentality}
\setcounter{equation}{0}
\label{sectrans}

One may associate with each term in an operator or amplitude 
a degree of transcendentality as follows:
each factor of $\zeta_k$, $\pi^k$, $\log^k z$
(where $z$ is any ratio of momentum invariants),
or any polylogarithm of total degree $k$ has transcendentality $k$
and the transcendentality of a product of factors is additive. 
By uniform transcendentality $k_0$, we mean that 
in an $\ep$-expansion $\sum_k a_k\ep^k$ (for dimensional regularization),
$a_k$ has transcendentality $k+k_0$. 
Maximal transcendentality means that $k_0$ has a maximal value.
In the case of ${\cal N}=4$ SYM and ${\cal N}=8$ supergravity, 
maximal transcendentality means that $L$-loop amplitudes
have uniform transcendentality $k_0 = 2L$;  
that is, all terms proportional to $(\lambda/\ep^2)^L\cdot 
\ep^k$ have degree of transcendentality $k$. 

In this section,
we first study the transcendentality 
of the IR-divergent Catani operators $\bI^\Ell$,
which determine all the IR-divergent terms of $n$-gluon amplitudes. 
The $\cN=4$ Catani operators have uniform transcendentality,
and moreover constitute the maximum transcendentality piece of the
corresponding QCD operators.  
We then go on to examine the transcendentality of
the subleading-color amplitudes of $\cN=4$ SYM theory.
These too have uniform transcendentality 
(as do the leading-color amplitudes),
but in this case do not constitute the entire
maximum transcendentality piece 
of the corresponding QCD amplitudes.

\subsection{Transcendentality of IR-divergent $n$-point operators}

{}From eqs.~(\ref{oneloopCatani})-(\ref{threeloopCatani}),
one may see that, for the $L$-loop amplitudes to have
maximal transcendentality $k_0= 2L$, the
$L$-loop Catani operator $\bI^\Ell$ must have 
uniform transcendentality $k_0 = 2 L$.
We will show 
in this subsection that the one- and two-loop Catani operators 
of $\cN=4$ SYM theory 
(and the three-loop operator, up to an undetermined term 
of $\cO(1/\ep)$) 
do satisfy this,
and moreover constitute the maximal transcendentality
piece of the QCD Catani operators. 
The review \cite{Bern:2004kq} is useful for this discussion. 

The one-loop Catani operator for QCD is \cite{Catani:1998bh}
\be
\bI^\One(\ep)
=\frac{1}{2}\sum_{i=1}^n\sum_{j\neq i}^n \bT_i\cdot \bT_j
\left[\frac{1}{\ep^2}
\left(\frac{\mu^2}{-s_{i,j}}\right)^\ep
+\frac{\gamma_i}{ \bT_i^2}\frac{1}{\ep}
\left(\frac{\mu^2}{-s_{i,j}}\right)^\ep
\right]
\label{Ioneqcd}
\ee
where $\gamma_i = b_0 = \frac{11}{6} N - \frac{2}{3} T_R N_f$ for
gluons.
The first term in brackets has uniform transcendentality $k_0=2$
while the second term in brackets has uniform transcendentality $k_0=1$,
and hence is subleading in transcendentality.
For ${\cal N}=4$ SYM, however, $b_0=0$, 
so in this case $\bI^\One(\ep)$ has uniform transcendentality $k_0=2$,
and is given by the maximal transcendentality part of 
the QCD operator (\ref{Ioneqcd}).

The two-loop Catani operator for QCD is \cite{Catani:1998bh}
\bea
\bI^{\Two} (\ep)
&=&-
\frac{1}{2}\bI^{(1)}(\ep)\left[\bI^\One(\ep)+\frac{2b_0}{\ep}\right]
+\cep\left[K + \frac{b_0}{\ep}\right]\bI^\One(2\ep)
\nonumber\\
&&+\frac{\cep}{4\ep}
\left[-\sum_{i=1}^n\sum_{j\neq i}^n
\bT_i\cdot \bT_j\frac{H_i^{\Two}}{\bT_i^2}
\left(\frac{\mu^2}{-s_{ij}}\right)^{2\ep}+\bhH^{\Two}(\ep)\right]
\eea
with $\bhH^\Two$ given by \cite{Bern:2004cz,MertAybat:2006wq,MertAybat:2006mz}
\be
\bhH^{\Two} =i\sum_{(i_1,i_2,i_3)}
f^{a_1a_2a_3}T_{i_1}^{a_1}T_{i_2}^{a_2}T_{i_3}^{a_3}
\log \left(\frac{ -s_{i_1i_2}}{-s_{i_2i_3}}\right)
\log \left(\frac{-s_{i_2i_3}}{-s_{i_3i_1}}\right)
\log \left(\frac{ -s_{i_3i_1}}{-s_{i_1i_2}}\right)\,.
\ee
Again, in general, $\bI^\Two$ contains terms of mixed 
transcendentality, 
but for ${\cal N}=4$ SYM, one has $b_0=0$, 
$K=-\zeta_2N$, and $H_i^{\Two}=\half\zeta_3N^2$ 
so only the terms of maximal transcendentality remain.
(The expression $\cep$ itself is of uniform 
transcendentality $k_0=0$, as can be seen from \eqn{cep}.) 
Thus we see that for ${\cal N}=4$ SYM, 
the operator $\bI^\Two$ has maximal 
uniform transcendentality $k_0=4$.

At three loops, we consider 
the ${\cal N}=4$ SYM Catani operator (\ref{Ithreeapprox}).
All the terms through $1/\ep^2$ 
have uniform transcendentality $k_0=6$, thus maximal. 
(The $1/\ep$ term contains $\bGam^\Three$, 
which we have not computed.)
Moreover, from eq. (30) of ref.~\cite{Sterman:2002qn},
we see that the additional terms in the three-loop QCD Catani operator, 
proportional to $b_0$,
contain at most terms of uniform transcendentality $k_0=5$, 
and thus subdominant. 
We expect the same pattern to hold for the 
three-loop anomalous dimension matrix $\bGam^{(3)}$. 

Based on these results, it is natural to expect that the 
$L$-loop $\cN=4$ SYM Catani operator $\bI^{(L)}$ will also be 
of uniform transcendentality,
and the maximal transcendentality piece of the QCD operator.

\subsection{Transcendentality of two-loop four-gluon amplitudes}

To examine the transcendentality of the $\cN=4$ SYM 
four-gluon amplitudes, one must look at the exact expressions
for the planar and non-planar loop integrals.
{}From the explicit Laurent expansions given in 
refs.~\cite{Bern:2005iz,Tausk:1999vh},  one may see that,
while the one- and two-loop planar integrals 
(\ref{Monedef}) and (\ref{Mtwodef})
have uniform transcendentality $k_0=2$ and $k_0=4$ respectively,
the two-loop nonplanar integral (\ref{twoloopNP}) does not,
as it contains terms of subleading transcendentality.

The leading-color one- and two-loop amplitudes  $A^\OneLC$ and $A^\TwoLC$,
which are built from one- and two-loop planar integrals
therefore have uniform transcendentality, as is already well known.
The subleading-color one-loop amplitude $A^\OneDT$
(and therefore the one-loop supergravity amplitude), 
which by \eqn{oneloopnp} is also built from the  one-loop planar integral,
also has uniform transcendentality $k_0=2$. 

It is not obvious that the two-loop subleading-color amplitudes
$A^\TwoDT$ and $A^\TwoSC$, 
which are built from two-loop planar and non-planar integrals
(cf. \eqns{twoloopDT} {twoloopSC})
have uniform transcendentality,
but we have verified,  
using the expressions in ref.~\cite{Naculich:2008ew},
that all the terms of subdominant transcendentality cancel
out, and that the full nonplanar amplitudes have 
uniform transcendentality $k_0=4$,
at least through ${\cal O}(\ep^0)$.

It was previously observed
that the two-loop $\cN=8$ supergravity amplitude (\ref{twoloopSG}),
which is also built from the two-loop non-planar integral (\ref{twoloopNP}),
nonetheless 
has uniform transcendentality \cite{Naculich:2008ew,Brandhuber:2008tf}.

Given that the $\cN=4$ SYM four-gluon amplitudes 
(at least through two loops) have uniform transcendentality,
the question arises whether they constitute the entire 
maximum transcendentality piece of the 
corresponding pure QCD amplitudes \cite{Kotikov:2004er}.

At one loop, the leading-color four-gluon QCD amplitude 
is \cite{Bern:2002tk}
\be
M^{\rm gluon}_{\lambda_1\lambda_2\lambda_3\lambda_4}
=(1-\ep\delta_R)M^{\rm scalar}_{\lambda_1\lambda_2\lambda_3\lambda_4}
-4M^{\cN=1}_{\lambda_1\lambda_2\lambda_3\lambda_4}
+M^{\cN=4}_{\lambda_1\lambda_2\lambda_3\lambda_4}
\label{maxtrans}
\ee
where $\lambda_i$ denote helicities.
The $\cN=4$ SYM four-gluon amplitude
$M^{\cN=4}_{\lambda_1\lambda_2\lambda_3\lambda_4}$
is nonzero only for helicities $--++$ or $-+-+$, 
and given by $Box^{(4)}(s,t)$,
which has uniform transcendentality, 
starting with $1/\ep^2$ 
(the $\ep^n$ term has transcendentality $n+2$).  
The terms 
$(1- \ep \delta_R) M^{\rm scalar}_{--++}$ and $M^{\cN=1}_{--++}$ 
are decomposed into 
$Bub^{(6)}$,
$\ep Box^{(8)}$,
$\ep Box^{(6)}$,
$Bub^{(4)}$, 
and terms with lower transcendentality, 
and from the explicit expressions in ref.~\cite{Bern:2002tk}, 
we see that they all start at most with $1/\ep$, 
thus at order $\ep^n$ have transcendentality at most $n+1$.
The terms
$(1- \ep \delta_R) M^{\rm scalar}_{-+-+}$ and $M^{\cN=1}_{-+-+}$,
however, contain the finite term $Box^{(6)}$,
which has pieces of (maximal) transcendentality two.
Hence, only in the case of the one-loop amplitude 
with helicity $--++$ are the maximal transcendentality terms 
of QCD given by the ${\cN}=4$ SYM result \cite{private}.

The one-loop U(1) decoupling identity (\ref{oneloopdecouple})
holds for both ${\cal N}=4$ SYM and for ${\cal N}=0$ (pure QCD). 
Since the maximal transcendentality terms of the 
leading-color one-loop four-gluon QCD amplitude with helicity $--++$
are given by the corresponding ${\cal N}=4$ SYM amplitudes,
the decoupling identity implies the 
same result for the subleading-color one-loop amplitudes. 

At two loops, 
this does not hold even for the leading-color amplitude with 
helicity $--++$.
The leading-color two-loop QCD amplitude \cite{Bern:2002tk} 
contains terms of transcendentality two that are not 
contained in the corresponding $\cN=4$ SYM amplitude\cite{private}.

\section{Conclusions}
\setcounter{equation}{0}
\label{secconcl}

In this paper we have studied the subleading-color 
(nonplanar) contributions to the $\cN=4$ SYM four-gluon amplitude.
Explicit expressions for the IR-divergent terms of the 
subleading-color amplitudes were computed through three loops
using the formalisms of Catani and of Sterman and Tejeda-Yeomans.
We extrapolated these results to conjecture the form 
of the leading IR divergences of the $N$-independent 
subleading-color amplitude $A^\EllL$  
in \eqns{Lodd}{Leven}.

We have also demonstrated some  connections between
$\cN=4$ SYM four-gluon amplitudes
and $\cN=8$ supergravity four-graviton amplitudes.  
The one-loop subleading-color SYM amplitude 
is proportional to the one-loop supergravity amplitude, 
the proportionality constant being the ratio of the
coupling constant $\lSYM$ for SYM
and the dimensionless effective coupling $\lSG u$ for supergravity.
Various relations exist between
the two-loop subleading-color SYM amplitudes and 
one- and two-loop supergravity amplitudes,
as detailed in sec.~\ref{secSG}.
The SYM/supergravity connection is most transparent in terms
of ratios of loop amplitudes to tree amplitudes $M^\Ell=A^\Ell/A^\Zero$.
The relation (\ref{llloop}) between $L$-loop SYM and $L$-loop supergravity 
amplitudes, which is valid for $L \le 2$, 
is understood by replacing $\lSYM$ with $\lSG u$ 
and summing over permutations.
The simple relation (\ref{llloop}), however,
fails at three loops and beyond 
(assuming that we have correctly determined the leading divergences 
of $\Msy^\EllL$ and $\MsgEll$).
Instead,  we obtain the relations (\ref{2k1}) and (\ref{2k2}), 
which do not have a simple interpretation. 
If \eqn{llloop} were correct 
(or \eqns{2k1}{2k2} had no extra terms), 
we would have had a simple, albeit nonintuitive, interpretation 
in terms of the 't Hooft picture 
(equating the topological expansion of SYM Feynman diagrams with
string worldsheets).  
Perhaps a modification of this picture can be found
that would relate the subleading-color (nonplanar) $\cN=4$ SYM amplitudes
to the $\cN=8$ supergravity amplitudes to all loop orders.

Our one and two-loop results suggest the possibility of 
a weak-weak duality between $\cN=4$ SYM and $\cN=8$ supergravity
(see also ref.~\cite{ArkaniHamed:2008gz}), 
in contrast to the usual strong-weak AdS/CFT duality. 
Such a duality, however,  would require a relation
between SYM and supergravity amplitudes at three loops and beyond,
a relation we have failed to find.  
Since the gauge theory expansion has two parameters, 
$\lSYM = g^2 N$ and $1/N$ 
(corresponding to $\alpha'$ and $g_s$ in the string picture),
whereas the loop expansion of $\cN=8$ supergravity has only one parameter, 
$\lSG = (\kappa/2)^2$,
it is perhaps unlikely that such a duality could exist
without taking into account stringy corrections.
It is possible that one needs to consider the mixing of
other string theory modes into the loop expansion of supergravity,
giving the extra terms in \eqns{2k1}{2k2}.

The one-loop supergravity amplitude appears in many different places 
in subleading-color SYM amplitudes.  
Examples include the two leading IR-divergent terms
in eqs.~(\ref{twoloopmat}), (\ref{threeeps}), (\ref{Lodd}),
and (\ref{Leven}), as well as the full IR divergence in 
eqs.~(\ref{IRSC}) and (\ref{IR21}).
                                                              
Finally, we have examined the issue of transcendentality of the
$\cN=4$ SYM subleading-color amplitudes and of the Catani operators.
Up to two loops,  the nonplanar amplitudes have uniform
transcendentality, as is already known for the planar amplitudes.
The $\cN=4$ SYM  Catani operators (at least through three loops)
also have uniform transcendentality,
and constitute the maximum transcendentality piece of the
QCD Catani operators. 

\vspace{.2in}
{\bf Acknowledgments} 
The authors would like to thank Fernando Alday for useful correspondence, 
Lance Dixon for correspondence and conversations,
and Radu Roiban for discussions.
We also thank Lance Dixon for alerting us 
to some inaccurate statements in sec.~\ref{sectrans} of {\tt v1} of this paper.
HN's research  has been done with partial support 
from  MEXT's program ``Promotion of Environmental Improvement for 
Independence of Young Researchers" under the Special Coordination 
Funds for Promoting Science and Technology, and also with partial support from MEXT KAKENHI grant nr. 20740128. 

\providecommand{\href}[2]{#2}\begingroup\raggedright\endgroup

\begin{thebibliography}{10}

\bibitem{Anastasiou:2003kj}
C.~Anastasiou, Z.~Bern, L.~J. Dixon, and D.~A. Kosower, ``{Planar amplitudes in
  maximally supersymmetric Yang-Mills theory},''
  \href{http://dx.doi.org/10.1103/PhysRevLett.91.251602}{{\em Phys. Rev. Lett.}
  {\bf 91} (2003)  251602},
\href{http://arxiv.org/abs/hep-th/0309040}{{\tt arXiv:hep-th/0309040}}.

\bibitem{Maldacena:1997re}
J.~M. Maldacena, ``{The large $N$ limit of superconformal field theories and
  supergravity},'' {\em Adv. Theor. Math. Phys.} {\bf 2} (1998)  231--252,
\href{http://arxiv.org/abs/hep-th/9711200}{{\tt arXiv:hep-th/9711200}}.

\bibitem{Gubser:1998bc}
S.~S. Gubser, I.~R. Klebanov, and A.~M. Polyakov, ``{Gauge theory correlators
  from non-critical string theory},''
  \href{http://dx.doi.org/10.1016/S0370-2693(98)00377-3}{{\em Phys. Lett.} {\bf
  B428} (1998)  105--114},
\href{http://arxiv.org/abs/hep-th/9802109}{{\tt arXiv:hep-th/9802109}}.

\bibitem{Aharony:1999ti}
O.~Aharony, S.~S. Gubser, J.~M. Maldacena, H.~Ooguri, and Y.~Oz, ``{Large $N$
  field theories, string theory and gravity},''
  \href{http://dx.doi.org/10.1016/S0370-1573(99)00083-6}{{\em Phys. Rept.} {\bf
  323} (2000)  183--386},
\href{http://arxiv.org/abs/hep-th/9905111}{{\tt arXiv:hep-th/9905111}}.

\bibitem{Bern:1997nh}
Z.~Bern, J.~S. Rozowsky, and B.~Yan, ``{Two-loop four-gluon amplitudes in
  ${\cal N}=4$ super-Yang- Mills},''
  \href{http://dx.doi.org/10.1016/S0370-2693(97)00413-9}{{\em Phys. Lett.} {\bf
  B401} (1997)  273--282},
\href{http://arxiv.org/abs/hep-ph/9702424}{{\tt arXiv:hep-ph/9702424}}.

\bibitem{Bern:1998ug}
Z.~Bern, L.~J. Dixon, D.~C. Dunbar, M.~Perelstein, and J.~S. Rozowsky, ``{On
  the relationship between Yang-Mills theory and gravity and its implication
  for ultraviolet divergences},''
  \href{http://dx.doi.org/10.1016/S0550-3213(98)00420-9}{{\em Nucl. Phys.} {\bf
  B530} (1998)  401--456},
\href{http://arxiv.org/abs/hep-th/9802162}{{\tt arXiv:hep-th/9802162}}.

\bibitem{Smirnov:1999gc}
V.~A. Smirnov, ``{Analytical result for dimensionally regularized massless
  on-shell double box},''
  \href{http://dx.doi.org/10.1016/S0370-2693(99)00777-7}{{\em Phys. Lett.} {\bf
  B460} (1999)  397--404},
\href{http://arxiv.org/abs/hep-ph/9905323}{{\tt arXiv:hep-ph/9905323}}.

\bibitem{Tausk:1999vh}
J.~B. Tausk, ``{Non-planar massless two-loop Feynman diagrams with four
  on-shell legs},'' \href{http://dx.doi.org/10.1016/S0370-2693(99)01277-0}{{\em
  Phys. Lett.} {\bf B469} (1999)  225--234},
\href{http://arxiv.org/abs/hep-ph/9909506}{{\tt arXiv:hep-ph/9909506}}.

\bibitem{Mueller:1979ih}
A.~H. Mueller, ``{On the asymptotic behavior of the Sudakov form-factor},''
\href{http://dx.doi.org/10.1103/PhysRevD.20.2037}{{\em Phys. Rev.} {\bf D20}
  (1979)  2037}.

\bibitem{Collins:1980ih}
J.~C. Collins, ``{Algorithm to compute corrections to the Sudakov form-factor},''
\href{http://dx.doi.org/10.1103/PhysRevD.22.1478}{{\em Phys. Rev.} {\bf D22}
  (1980)  1478}.

\bibitem{Collins:1989bt}
J.~C. Collins, ``Sudakov form factors,'' {\em Adv. Ser. Direct. High Energy
  Phys.} {\bf 5} (1989)  573--614,
\href{http://arxiv.org/abs/hep-ph/0312336}{{\tt hep-ph/0312336}}.

\bibitem{Sen:1981sd}
A.~Sen, ``{Asymptotic Behavior of the Sudakov Form-Factor in QCD},''
\href{http://dx.doi.org/10.1103/PhysRevD.24.3281}{{\em Phys. Rev.} {\bf D24}
  (1981)  3281}.

\bibitem{Magnea:1990zb}
L.~Magnea and G.~Sterman, ``Analytic continuation of the Sudakov form-factor in
  QCD,''
{\em Phys. Rev.} {\bf D42} (1990)  4222--4227.

\bibitem{Giele:1991vf}
W.~T. Giele and E.~W.~N. Glover, ``{Higher order corrections to jet
  cross-sections in $e^+ e^-$ annihilation},''
\href{http://dx.doi.org/10.1103/PhysRevD.46.1980}{{\em Phys. Rev.} {\bf D46}
  (1992)  1980--2010}.

\bibitem{Kunszt:1994mc}
Z.~Kunszt, A.~Signer, and Z.~Trocsanyi, ``{Singular terms of helicity
  amplitudes at one loop in QCD and the soft limit of the cross-sections of
  multiparton processes},''
  \href{http://dx.doi.org/10.1016/0550-3213(94)90077-9}{{\em Nucl. Phys.} {\bf
  B420} (1994)  550--564},
\href{http://arxiv.org/abs/hep-ph/9401294}{{\tt arXiv:hep-ph/9401294}}.

\bibitem{Catani:1996vz}
S.~Catani and M.~H. Seymour, ``{A general algorithm for calculating jet cross
  sections in NLO QCD},''
  \href{http://dx.doi.org/10.1016/S0550-3213(96)00589-5}{{\em Nucl. Phys.} {\bf
  B485} (1997)  291--419},
\href{http://arxiv.org/abs/hep-ph/9605323}{{\tt arXiv:hep-ph/9605323}}.

\bibitem{Dixon:2008gr}
L.~J. Dixon, L.~Magnea, and G.~Sterman, ``{Universal structure of subleading
  infrared poles in gauge theory amplitudes},''
  \href{http://dx.doi.org/10.1088/1126-6708/2008/08/022}{{\em JHEP} {\bf 08}
  (2008)  022},
\href{http://arxiv.org/abs/0805.3515}{{\tt arXiv:0805.3515 [hep-ph]}}.

\bibitem{Catani:1998bh}
S.~Catani, ``The singular behaviour of {QCD} amplitudes at two-loop order,''
  {\em Phys. Lett.} {\bf B427} (1998)  161--171,
\href{http://arxiv.org/abs/hep-ph/9802439}{{\tt hep-ph/9802439}}.

\bibitem{Sterman:2002qn}
G.~Sterman and M.~E. Tejeda-Yeomans, ``Multi-loop amplitudes and resummation,''
  {\em Phys. Lett.} {\bf B552} (2003)  48--56,
\href{http://arxiv.org/abs/hep-ph/0210130}{{\tt hep-ph/0210130}}.

\bibitem{Bern:2005iz}
Z.~Bern, L.~J. Dixon, and V.~A. Smirnov, ``{Iteration of planar amplitudes in
  maximally supersymmetric Yang-Mills theory at three loops and beyond},''
  \href{http://dx.doi.org/10.1103/PhysRevD.72.085001}{{\em Phys. Rev.} {\bf
  D72} (2005)  085001},
\href{http://arxiv.org/abs/hep-th/0505205}{{\tt arXiv:hep-th/0505205}}.

\bibitem{'tHooft:1973jz}
G.~'t~Hooft, ``{A planar diagram theory for strong interactions},''
\href{http://dx.doi.org/10.1016/0550-3213(74)90154-0}{{\em Nucl. Phys.} {\bf
  B72} (1974)  461}.

\bibitem{Naculich:2008ew}
S.~G. Naculich, H.~Nastase, and H.~J. Schnitzer, ``{Two-loop graviton
  scattering relation and IR behavior in ${\cal N}=8$ supergravity},''
{{\em Nucl. Phys.} {\bf B805} (2008)  40--58},
\href{http://arxiv.org/abs/0805.2347}{{\tt arXiv:0805.2347v3 [hep-th]}}.

\bibitem{Brandhuber:2008tf}
A.~Brandhuber, P.~Heslop, A.~Nasti, B.~Spence, and G.~Travaglini, ``{Four-point
  Amplitudes in ${\cal N}=8$ Supergravity and Wilson Loops},''
\href{http://arxiv.org/abs/0805.2763}{{\tt arXiv:0805.2763 [hep-th]}}.

\bibitem{Bern:1990ux}
Z.~Bern and D.~A. Kosower, ``{Color decomposition of one loop amplitudes in
  gauge theories},''
\href{http://dx.doi.org/10.1016/0550-3213(91)90567-H}{{\em Nucl. Phys.} {\bf
  B362} (1991)  389--448}.

\bibitem{Glover:2001af}
E.~W.~N. Glover, C.~Oleari, and M.~E. Tejeda-Yeomans, ``{Two-loop QCD
  corrections to gluon gluon scattering},''
  \href{http://dx.doi.org/10.1016/S0550-3213(01)00210-3}{{\em Nucl. Phys.} {\bf
  B605} (2001)  467--485},
\href{http://arxiv.org/abs/hep-ph/0102201}{{\tt arXiv:hep-ph/0102201}}.

\bibitem{Green:1987sp}
M.~B. Green, J.~H. Schwarz, and E.~Witten, {\it Superstring Theory}, vol. 1
  (Cambridge University Press, 1987,  Cambridge).

\bibitem{Green:1982sw}
M.~B. Green, J.~H. Schwarz, and L.~Brink, ``{${\cal N}=4$ Yang-Mills and ${\cal
  N}=8$ Supergravity as Limits of String Theories},''
\href{http://dx.doi.org/10.1016/0550-3213(82)90336-4}{{\em Nucl. Phys.} {\bf
  B198} (1982)  474--492}.

\bibitem{Bern:2002tk}
Z.~Bern, A.~De~Freitas, and L.~J. Dixon, ``{Two-loop helicity amplitudes for
  gluon gluon scattering in QCD and supersymmetric Yang-Mills theory},'' {\em
  JHEP} {\bf 03} (2002)  018,
\href{http://arxiv.org/abs/hep-ph/0201161}{{\tt arXiv:hep-ph/0201161}}.

\bibitem{Bern:2004cz}
Z.~Bern, L.~J. Dixon, and D.~A. Kosower, ``{Two-loop g $\to$ g g splitting
  amplitudes in QCD},'' {\em JHEP} {\bf 08} (2004)  012,
\href{http://arxiv.org/abs/hep-ph/0404293}{{\tt arXiv:hep-ph/0404293}}.

\bibitem{MertAybat:2006wq}
S.~Mert~Aybat, L.~J. Dixon, and G.~Sterman, ``{The two-loop anomalous dimension
  matrix for soft gluon exchange},''
  \href{http://dx.doi.org/10.1103/PhysRevLett.97.072001}{{\em Phys. Rev. Lett.}
  {\bf 97} (2006)  072001},
\href{http://arxiv.org/abs/hep-ph/0606254}{{\tt arXiv:hep-ph/0606254}}.

\bibitem{MertAybat:2006mz}
S.~Mert~Aybat, L.~J. Dixon, and G.~Sterman, ``{The two-loop soft anomalous
  dimension matrix and resummation at next-to-next-to leading pole},''
  \href{http://dx.doi.org/10.1103/PhysRevD.74.074004}{{\em Phys. Rev.} {\bf
  D74} (2006)  074004},
\href{http://arxiv.org/abs/hep-ph/0607309}{{\tt arXiv:hep-ph/0607309}}.


\bibitem{Kawai:1985xq}
H.~Kawai, D.~C. Lewellen, and S.~H.~H. Tye, ``{A Relation Between Tree
  Amplitudes of Closed and Open Strings},''
\href{http://dx.doi.org/10.1016/0550-3213(86)90362-7}{{\em Nucl. Phys.} {\bf
  B269} (1986)  1}.


\bibitem{ArkaniHamed:2008gz}
N.~Arkani-Hamed, F.~Cachazo, and J.~Kaplan, ``{What is the Simplest Quantum
  Field Theory?}''
\href{http://arxiv.org/abs/0808.1446}{{\tt arXiv:0808.1446 [hep-th]}}.

\bibitem{Bern:2004kq}
Z.~Bern, L.~J. Dixon, and D.~A. Kosower, ``{${\cal N}=4$ super-Yang-Mills
  theory, QCD and collider physics},''
  \href{http://dx.doi.org/10.1016/j.crhy.2004.09.007}{{\em Comptes Rendus
  Physique} {\bf 5} (2004)  955--964},
\href{http://arxiv.org/abs/hep-th/0410021}{{\tt arXiv:hep-th/0410021}}.

\bibitem{Kotikov:2004er}
A.~V. Kotikov, L.~N. Lipatov, A.~I. Onishchenko, and V.~N. Velizhanin,
``Three-loop universal anomalous dimension of the Wilson operators in
${\cal N} = 4 $ SUSY Yang-Mills model,'' {\em Phys. Lett.} {\bf B595} (2004)
  521--529,
\href{http://arxiv.org/abs/hep-th/0404092}{{\tt hep-th/0404092}}.

\bibitem{private}
L.~J.~Dixon, private communication.

\end{thebibliography}
\end{document}